\documentclass[11pt]{article}
\usepackage[margin=1in]{geometry}
\usepackage{graphicx}
\usepackage{authblk}
\usepackage{setspace}
\usepackage{url}
\usepackage{multirow}
\usepackage{multicol}
\usepackage[T1]{fontenc}
\usepackage[usenames, dvipsnames]{xcolor}
\usepackage{bm}
\usepackage{amsmath}
\usepackage{amssymb}
\usepackage{amsthm}
\usepackage{mathtools}
\usepackage{bbm}
\usepackage{commath}
\usepackage[authoryear, sort, round]{natbib}
\usepackage[bottom]{footmisc}

\DeclareMathOperator*{\argmax}{arg\,max}
\newcommand{\R}{\mathbb{R}}

\def\P{\mathbb{P}}

\newcommand{\ind}[1]{\mathbf{1}\big\{#1\big\}} 

\newcommand{\x}{\bm{X}}

\newcommand{\bmb}{\bm{\beta}}

\DeclarePairedDelimiterX{\inp}[2]{\langle}{\rangle}{#1, #2}

\theoremstyle{plain}
\newtheorem{theorem}{Theorem}[section]

\newtheorem{proposition}[theorem]{Proposition}

\theoremstyle{definition}
\newtheorem{definition}[theorem]{Definition}

\begin{document}

\title{\textbf{Bayesian changepoint detection via logistic regression \\ and the topological analysis of image series}}
\author[1]{Andrew M. Thomas}
\author[2]{Michael Jauch} 
\author[3]{David S. Matteson}
\affil[1]{Department of Statistics and Actuarial Science, University of Iowa}
\affil[2]{Department of Statistics, Florida State University}
\affil[3]{Department of Statistics and Data Science, Cornell University}
\date{}

\maketitle

\begin{abstract}

We present a Bayesian method for multivariate changepoint detection that allows for simultaneous inference on the location of a changepoint and the coefficients of a logistic regression model for distinguishing pre-changepoint data from post-changepoint data. In contrast to many methods for multivariate changepoint detection, the proposed method is applicable to data of mixed type and avoids strict assumptions regarding the distribution of the data and the nature of the change. The regression coefficients provide an interpretable description of a potentially complex change. For posterior inference, the model admits a simple Gibbs sampling algorithm based on Pólya-gamma data augmentation. We establish conditions under which the proposed method is guaranteed to recover the true underlying changepoint. As a testing ground for our method, we consider the problem of detecting topological changes in time series of images. We demonstrate that our proposed method BCLR, combined with a topological feature embedding, performs well on both simulated and real image data.
The method also successfully recovers the location and nature of changes in more traditional changepoint tasks. 

\end{abstract}

\noindent%

\section{Introduction}

Time series often consist of homogeneous segments interrupted by abrupt structural changes. Changepoint analysis involves determining the number, locations, and nature of these \emph{changepoints}. Statistical methods for changepoint analysis have a long history, with notable early work by \citeauthor{page1954} \citeyearpar{page1954, page1955}. Since then, many parametric models (see \citet{chen2012} for an overview) 
and nonparametric approaches \citep{bhattacharya1968, brodsky1993} have been proposed. Changepoint methods have been applied in diverse fields such as global finance \citep{allen2018}, climatology \citep{balaji2018}, bioinformatics \citep{fan_mackey, liu2018}, dairy science \citep{lombard2020}, hydrology \citep{water}, and hygiene \citep{hygiene}.



Changepoint analysis, especially in the multivariate setting, is a hard problem. We highlight a few challenges that motivate the present article: 

\vspace{5pt}
\noindent \textbf{The model for the data.} Conventional likelihood-based approaches to changepoint analysis require the specification of a model for the data within each homogeneous segment of the observed time series. There is a tradeoff between fidelity to the data and parsimony of the model (which is typically closely related to its computational tractability). On one end of the spectrum, there are simple parametric methods that make restrictive assumptions regarding the distribution of the data within each segment, e.g. that the data are Gaussian \citep{srivastava1986, lavielle2006} or follow an exponential family distribution \citep{chen2012}. On the other end, there are elaborate Bayesian nonparametric methods that avoid restrictive assumptions but may be difficult for the non-expert to implement or interpret \citep{martinez_mena, corradin2022}. 
Negotiating this tradeoff between fidelity and parsimony becomes much more difficult in the context of multivariate time series, and even more so when the time series include both continuous and discrete components.

\vspace{5pt}
\noindent \textbf{The nature of the changepoint.} Statistical methods for changepoint analysis differ in the assumptions they make regarding the nature of the changepoints. Methods developed to detect simple changes (e.g. in mean or covariance, as in \citealp{lavielle2006,  jin2022bayesian}) 
may miss the complex changes that can occur in multivariate time series, while methods developed to detect arbitrary changes \citep[for example]{matteson2014, arlot2019} may lack power or lead to results that are hard to interpret.

\vspace{5pt}
\noindent \textbf{Uncertainty quantification.} Changepoint analysis requires making several related inferences regarding the number, locations, and nature of the changepoints. Developing methods that propagate uncertainty across these inferences is an essential yet challenging task. Bayesian approaches are a natural solution, and a variety of Bayesian methods have been developed to quantify uncertainty on the location and number of the changepoints \citep{carlin1992, barry_hartigan, loschi_cruz, fan_mackey, bardwell2017, quinlan2022}. Quasi- or generalized Bayesian approaches  to changepoint detection---such as \citet{casini2022}---also provide uncertainty quantification for the location of changepoints. 
\vspace{5pt}


To address these challenges, we introduce a new method for Bayesian changepoint analysis in the offline setting called BCLR. Though the method we devise is tailored to the setting where there is at most one changepoint, we detail an extension to the multiple changepoint setting in Section 6 and present results of the multiple changepoint method in Section S6 of the Supplementary Material. Our method allows for simultaneous inference on the location of a changepoint and the coefficients of a logistic regression model for distinguishing pre-changepoint data from post-changepoint data. The regression coefficients provide an interpretable description of a potentially complex change. Because the observed time series is treated as a sequence of covariate vectors, there is no need to specify a model for the data, and the method can be applied to data of mixed type. 
For posterior inference, the model admits a simple Gibbs sampling algorithm based on Pólya-gamma data augmentation \citep{polson2013}. We establish conditions under which the proposed method is guaranteed to recover the true underlying changepoint.


Several other recent articles have explored the idea of using classifiers for changepoint detection. \citet{londschien2022} leverage the class probability predictions from a classifier (e.g. a random forest) to construct a classifier log-likelihood ratio that can be used to compare potential changepoint configurations. \citet{constrastive_CP} shares a similar spirit but focuses on the online setting. In a different direction, \citet{li2023} use neural networks and labeled examples of changepoints to construct new test statistics for detecting changes. 

There are two important differences between our proposed method and the methods presented in these articles. The first is that our method 
leverages classification for changepoint analysis within a Bayesian framework. 
As a result, we are able to incorporate prior information into our analysis, to quantify uncertainty related to the unknown parameters, and to take advantage of an extensive collection of computational techniques for posterior inference. 
The proposed Bayesian formulation is nonstandard in the sense that the posterior results from conditioning on the event that a collection of binary response variables has a changepoint structure rather than conditioning on the precise values, which are unobserved. It can also be viewed as a special case of a more flexible quasi- or generalized Bayesian formulation \citep{cherno2003, bissiri2016}. 
The second distinguishing feature of our approach is that the regression coefficients estimated with our method provide an interpretable description of a potentially complex change. 
Methods based on random forests and neural networks are harder to interpret.


As a testing ground for our method, we consider the problem of detecting topological changes in time series of images. Most methods in the image change detection literature consider pixelwise differences or small sequences of images \citep{imc_survey}. 
These methods fail when faced with substantial noise. One can develop more robust methods by focusing on those quantitative summaries or \textit{features} 
of an image series most relevant to detecting a change. Topological data analysis (TDA) has gained traction in the statistics and machine learning communities by developing features that lead to improved classification \citep{top_ml}. For example, \citet{effective_ph} and \citet{obayashi2018} demonstrate that TDA, and persistent homology in particular, is effective for learning various nonlinear features of image data in an off-the-shelf fashion. 
We demonstrate that the proposed changepoint method, in conjunction with a topological feature embedding, successfully recovers the location and topological nature of nonlinear changes in image series.

We now outline the remainder of the article. In Section 2, we introduce the proposed changepoint model and provide theoretical results that shed light on its efficacy. Section 3 reviews important concepts from topological data analysis and describes the feature embedding we use for detecting topological changes in image series. 
In Section 4, we evaluate the proposed model on simulated image data and in two other important changepoint settings. The first involves data of mixed type, while the second involves a change in covariance. We find that the performance of our method is comparable to or better than that of the state-of-the-art methods and that our method provides useful information not available from competing methods. In Section 5, we evaluate the method on real-world image data, and the conclusions we see are consistent with those of Section 4. 
Section 6 details an extension to the multiple changepoint setting which  
retains the fundamental features of the single changepoint method as well as its 
competitive performance.
We conclude in Section 7 with a summary of our main contributions and a discussion of future directions. 


\section{Bayesian changepoint detection via logistic regression}\label{s:bclr2}

We begin, in the first subsection, by introducing the proposed changepoint method for the single changepoint setting. In the second subsection, we present two important theoretical properties that help illuminate how the method works.


\subsection{The changepoint model} \label{ss:bcp}

The proposed changepoint method can be understood from two complementary perspectives: a Bayesian perspective and a quasi-Bayesian perspective. The Bayesian formulation is nonstandard because the posterior results from conditioning on an event other than the usual observation of data. This nonstandard formulation retains the advantages of a Bayesian approach without requiring us to specify a model for the data within each homogeneous segment of the observed time series. In this sense, our method has something in common with \citet{hoff07}, \citet{Miller19}, and \citet{lewis21}, each of which propose to overcome some challenge associated with a conventional Bayesian analysis by conditioning on a carefully chosen event. The quasi-Bayesian formulation shares these same advantages but offers more flexibility to incorporate prior information regarding the changepoint. Quasi-Bayesian methods, also known as generalized Bayesian methods \citep{bissiri2016}, are well-established in the statistics and econometrics literatures, and are typically motivated as a robust, nonparametric alternative to classical or Bayesian estimation. In the econometrics literature, quasi-Bayesian methods are also referred to as Laplace-type methods \citep{cherno2003, casini2022}. 

\subsubsection{The Bayesian formulation}

Let $\bm{x}_1, \ldots, \bm{x}_n \in \mathbbm{R}^d$ be a multivariate time series that we expect to have a single changepoint. In many cases, the series $\bm{x}_1, \ldots, \bm{x}_n$ is constructed from a raw series $\tilde{\bm{x}}_1, \ldots, \tilde{\bm{x}}_n$ using a feature mapping $\psi: \tilde{\bm{x}}_i \mapsto \bm{x}_i$ chosen to better represent the change of interest. We suppose there is a latent variable $Y_i \in \{0,1\}$ associated with each $\bm{x}_i$ and  that 
\begin{align} \label{dist_for_y}
Y_i \mid \bm{\beta}, \bm{x}_i \stackrel{\text{ind.}}{\sim} \text{Bernoulli}\left(\frac{e^{\bm{x}_i^\top \bm{\beta}}}{1 + e^{\bm{x}_i^\top \bm{\beta}}}\right),
\end{align} 
where $\bm{\beta} \in \mathbbm{R}^d$ is a vector of unknown regression coefficients. Taking a Bayesian perspective, we assign $\bm{\beta}$ a Gaussian prior, $\bm{\beta} \sim \text{N}\left(\bm{\mu}, \Sigma\right).$ Letting $\bm{Y} = (Y_1, \ldots, Y_n)^\top$ and $\bm{X}$ be the $n \times d$ matrix whose $i^{\text{th}}$ row is $\bm{x}_i^\top,$ the prior distribution for $\bm{\beta}$ and the conditional distribution specified in \eqref{dist_for_y} define a joint distribution for $\left(\bm{Y}, \bm{\beta}\right)$ with a density $p(\bm{y}, \bm{\beta} \mid \bm{X}).$ 
In the standard logistic regression setting, we would condition on the precise value of the response vector $\bm{Y} = \bm{y}$ to get the posterior density $p\left(\bm{\beta} \mid \bm{Y} = \bm{y}, \bm{X}\right).$ Tuning-free posterior inference could then be carried out with P{\'o}lya-Gamma data augmentation as described in \cite{polson2013}. 

In our setting, we do not observe the precise value of $\bm{Y},$ but we know that the times series $\bm{x}_1, \ldots, \bm{x}_n$ has a single changepoint $\kappa \in \{1, \ldots, n-1\}.$ We can condition on this changepoint structure as follows. Let $\Gamma_n$ be the set of binary vectors of length $n$ such that the first $\kappa$ entries are zeros and the last $n-\kappa$ entries are ones. 
Conditioning on the event $\bm{Y} \in \Gamma_n$ leads to a posterior distribution for $\left(\bm{Y}, \bm{\beta}\right)$ with density $p\left(\bm{y}, \bm{\beta} \mid \bm{Y} \in \Gamma_n, \bm{X}\right) \propto p(\bm{y}, \bm{\beta} \mid \bm{X}) \mathbf{1}\{\bm{y} \in \Gamma_n\}.$ Because there is a one-to-one correspondence between elements of $\Gamma_n$ and locations of the changepoint $\kappa,$ a simple change of variables leads to a posterior distribution over $\left(\kappa,\bm{\beta}\right).$
The posterior density for $\left(\kappa, \bm{\beta}\right),$ which we denote by $\pi_\text{B}\left(\kappa, \bm{\beta}\mid \bm{X}\right)$, satisfies
\begin{align} \label{posterior_kappa_beta}
\pi_\text{B}\left(\kappa, \bm{\beta} \mid \bm{X}\right) 
&\propto \left\{\prod_{i=1}^n\left(\frac{1}{1+e^{\bm{x}_i^\top \bm{\beta}}}\right)^{\mathbf{1}\{i \leq \kappa\}} \left(\frac{e^{\bm{x}_i^\top \bm{\beta}}}{1+e^{\bm{x}_i^\top \bm{\beta}}}\right)^{\mathbf{1}\{i > \kappa\}}\right\} \, \pi_\text{B}\left(\bm{\beta}\right) \nonumber \\ 
&=\left\{\prod_{i=1}^\kappa \frac{1}{1+e^{\bm{x}_i^\top \bm{\beta}}} \prod_{i=\kappa+1}^n \frac{e^{\bm{x}_i^\top \bm{\beta}}}{1+e^{\bm{x}_i^\top \bm{\beta}}}\right\}\, \pi_\text{B}\left(\bm{\beta}\right)  
\end{align} where $\pi_\text{B}\left(\bm{\beta}\right)$ is the multivariate normal prior density for $\bm{\beta}.$ 
Notice that we arrive at a posterior distribution over the changepoint $\kappa$ without explicitly specifying a prior distribution for it. 



We can derive a Gibbs sampler for posterior simulation by adapting the P{\'o}lya-Gamma data augmentation scheme of \citet{polson2013}. Data augmentation is necessary because there is no efficient way to directly simulate from the full conditional density $\pi_\text{B}\left(\bm{\beta}\mid \kappa, \bm{X}\right)$ arising from \eqref{posterior_kappa_beta}. The idea of data augmentation is to augment the parameter space by introducing additional latent variables such that 1) we recover the original posterior distribution when we marginalize over the latent variables and 2) we can easily simulate from the resulting full conditional distributions. The full conditional density $\pi_\text{B}\left(\bm{\beta}\mid \kappa, \bm{X}\right)$ is identical to the posterior density of the regression coefficients in a logistic regression model with a multivariate normal prior. Thus, we can leverage P{\'o}lya-Gamma data augmentation, which has become a standard approach to posterior simulation in Bayesian methods related to logistic regression. We augment the parameter space with a vector $\bm{\omega} = \left(\omega_1, \ldots, \omega_n\right)$ of  P{\'o}lya-Gamma latent variables, yielding a posterior density over the parameters $\left(\kappa, \bmb, \bm{\omega}\right)$ with tractable full conditional distributions. The full conditional distributions for $\bmb$ and the $\omega_i$'s are identical to those appearing in  \cite{polson2013}. The full conditional distribution for $\kappa$ is a discrete distribution supported on the set $\{1, \ldots, n-1\}$ and satisfies 
\begin{equation} \label{e:kbetax}
\pi_{\text{B}}\left(\kappa \mid \bm{\beta}, \bm{X}\right) \propto \prod_{i=1}^\kappa \frac{1}{1+e^{\bm{x}_i^\top \bmb}} \prod_{i=\kappa+1}^n \frac{e^{\bm{x}_i^\top \bmb}}{1+e^{\bm{x}_i^\top \bmb}}.
\end{equation} 
Putting this all together, the Gibbs sampler iterates through the following steps: 
\begin{align*}
\kappa \mid \bm{\beta}, \bm{X} &\sim \pi_{\text{B}}\left(\cdot\mid \bm{\beta}, \bm{X}\right) \\ 
\omega_i \mid \bm{\beta} &\stackrel{\text{ind.}}{\sim} \text{PG}\left(1, \bm{x}_i^\top\bm{\beta}\right) \\ 
\bm{\beta} \mid \kappa, \bm{\omega} &\sim \text{N}\left(\bm{m}_\omega, \bm{V}_\omega\right),
\end{align*}
where 
\begin{align*}
\bm{V}_\omega &= \left(\bm{X}^\top\Omega \bm{X} + \Sigma^{-1}\right)^{-1} \\ 
\bm{m}_\omega &= \bm{V}_\omega\left(\bm{X}^\top\bm{\delta} + \Sigma^{-1}\bm{\mu}\right)
\end{align*}
and $\Omega = \text{diag}\left(\bm{\omega}\right).$ The first $\kappa$ entries of the vector $\bm{\delta}$ are equal to $-1/2$ while the last $n-\kappa$ entries are equal to $1/2.$

\subsubsection{The quasi-Bayesian formulation}

The proposed changepoint method can also be understood as quasi-Bayesian in the sense of \cite{cherno2003}. 
In the quasi-Bayesian formulation, we directly define a \textit{quasi-likelihood}  
\begin{equation*}\label{e:lkl}
\mathcal{Q}(\bmb, \kappa \mid \bm{X})  =  \prod_{i=1}^{\kappa} \frac{1}{1+e^{\bm{x}_i^{\top}\bmb}} \prod_{i=\kappa+1}^{n} \frac{e^{\bm{x}_i^{\top}\bmb}}{1+e^{\bm{x}_i^{\top}\bmb}}
\end{equation*} 
that relates the parameters $\left(\kappa, \bmb\right)$ to the observations $\bm{x}_1, \ldots, \bm{x}_n.$ The \textit{quasi-posterior} density is then proportional to the product of the quasi-likelihood and the prior density $\pi_{\text{QB}}(\kappa, \bmb):$
$$
\pi_{\text{QB}}(\kappa, \bmb \mid \bm{X}) \propto \mathcal{Q}(\bmb, \kappa \mid \bm{X}) \pi_{\text{QB}}(\kappa, \bmb).
$$
The quasi-likelihood is bounded above by one, implying that a proper prior yields a proper quasi-posterior. If we suppose that $\pi_{\text{QB}}(\kappa, \bmb) = \pi_{\text{QB}}\left(\kappa\right)\pi_{\text{QB}}\left(\bmb\right),$ we can simulate from the quasi-posterior with a Gibbs sampler nearly identical to that of the previous section. 
The full conditional distribution for $\kappa$ becomes 
\begin{equation} \label{e:kquasi}
\pi_{\text{QB}}\left(\kappa \mid \bm{\beta}, \bm{X}\right) \propto \mathcal{Q}(\bmb, \kappa \mid \bm{X}) \pi_{\text{QB}}\left(\kappa\right)
\end{equation} while the full conditional distributions for $\bmb$ and the $\omega_i$'s remain the same. When the prior for $\kappa$ is uniform, the quasi-posterior is equivalent to the posterior of the previous section. Throughout the rest of the article, we will use a uniform prior for $\kappa$ unless otherwise noted. Thus, we drop the subscripts that distinguish the posterior from the quasi-posterior. 

In our experiments, we found that the posterior distribution tends to concentrate on regions of the parameter space with $\kappa = 1$ or $\kappa = n-1$ unless we omit the intercept and center the data. A discussion and justification of these preprocessing steps can be found in Section S1.1 of the Supplementary Material. In light of the empirical evidence and the exposition in Section S1.1, we henceforth center our data $\bm{x}_1, \dots, \bm{x}_n$ and omit the intercept from the linear term $\bm{x}_i^\top\bmb$. Although it is not strictly necessary, we also standardize the series to have sample variance 1, to permit the use of a default prior for $\bmb$ across data of different scales.

\subsection{Theoretical properties} \label{ss:theory}

Here we show that if the pre- and post-changepoint values of the linear functionals are sufficiently well-separated, our method will return the correct changepoint. Recall the definition of $\pi(\kappa \mid \bmb, \x)$ appears at \eqref{e:kbetax} (or at \eqref{e:kquasi} with $\pi(\kappa) \propto 1$). Let us denote the true changepoint as $\kappa^*$. For convenience of notation we will use $F$ to denote the inverse logit, where $F(x) := e^{x}/(1+e^x).$

\begin{proposition} \label{p:unimod}
If there exists some $\gamma > 0$ such that $\bm{x}_i^\top\bm{\beta} < - \gamma$ for $i \leq \kappa^*$ and $\bm{x}_i^\top\bm{\beta} > \gamma$ for $i > \kappa^*$, then 
\[
\argmax_{\kappa} \, \pi(\kappa \mid \bm{\beta}, \x) = \kappa^*
\]
and the probability mass function $\kappa \mapsto \pi(\kappa \mid \bm{\beta}, \bm{x})$ is unimodal. 
\end{proposition}
\begin{proof}
We begin by observing that for any possible $\kappa = 1, \dots, n-1$,
\[
\frac{\pi(\kappa+1 \mid \bm{\beta}, \x)}{\pi(\kappa \mid \bm{\beta}, \x)} = \frac{1-F(\bm{x}_{\kappa+1}^\top\bm{\beta})}{F(\bm{x}_{\kappa+1}^\top\bm{\beta})}.
\]
For $i \leq \kappa^*$ we have $ F(\bm{x}_{i}^\top\bm{\beta}) < F(-\gamma)$
and for $i > \kappa^*$ that $F(\bm{x}_{i}^\top\bm{\beta}) > F(\gamma)$.
Thus, the above yields for $\kappa < \kappa^*$ that
\[
\frac{\pi(\kappa+1 \mid \bm{\beta}, \x)}{\pi(\kappa \mid \bm{\beta}, \x)} > \frac{F(\gamma)}{1-F(\gamma)} > 1,
\]
and for $\kappa \geq \kappa^*$ that
\[
\frac{\pi(\kappa+1 \mid \bm{\beta}, \x)}{\pi(\kappa \mid \bm{\beta}, \x)} < \frac{1-F(\gamma)}{F(\gamma)} < 1,
\]
by symmetry of the logistic function at 0.
\end{proof}

Given that the data before and after the true changepoint $\kappa^* $ is well separated by some hyperplane, we will recover the changepoint. This leads to the following proposition that guarantees a large probability for said changepoint if a large margin is present. 

\begin{proposition}\label{c:m_bd}
If $\bmb \in A_{\kappa^*}^{\gamma}(\x)$ we have
\[
\pi(\kappa^* \mid \bmb, \x) > \frac{1-e^{-\gamma}}{1+e^{-\gamma}}.
\]
where $A_m^{\gamma}(\x) := \{\bm{\alpha} \in \R^d: \bm{x}_i^{\top}\bm{\alpha} < -\gamma, \, i \leq m \ \mathrm{and} \ \bm{x}_i^{\top}\bm{\alpha} > \gamma, \, i > m\}$. 
\end{proposition} 

\begin{proof}
Because $F(x) = e^x/(1+e^x)$ we have $F(\gamma)/[1-F(\gamma)]= e^\gamma$. Suppose without loss of generality that $\kappa^* = m$. Therefore if $\bm{\beta} \in A_m^{\gamma}(\x)$ we have by Proposition~\ref{p:unimod} that
\[
\pi(m \mid \bm{\beta}, \x) > e^\gamma \pi(m-1 \mid \bm{\beta}, \x),
\]
and similarly
\[
e^{-\gamma} \pi(m \mid \bm{\beta}, \x) > \pi(m+1 \mid \bm{\beta}, \x).
\]
Thus we conclude that 
\[
\pi(\kappa \mid \bm{\beta}, \x) < e^{-|\kappa-m|\gamma} \pi(m \mid \bm{\beta}, \x)
\]
Hence, we have 
\begin{align*}
1 &< \pi(m \mid \bm{\beta}, \x) \sum_{k=1}^n e^{-|k-m|\gamma} \\
&< \pi(m \mid \bm{\beta}, \x) \Big(1 + 2\sum_{k=1}^{\infty} e^{-k\gamma}\Big) \\
&= \pi(m \mid \bm{\beta}, \x) \big[1 + 2e^{-\gamma}/(1-e^{-\gamma})\big],
\end{align*}
which upon noticing that $1 + 2e^{-\gamma}/(1-e^{-\gamma}) = (1+e^{-\gamma})/(1-e^{-\gamma})$ finishes the proof.
\end{proof}

In Proposition S1.1 of Section S1.2 of the Supplementary Material, we prove an additional representation of the marginal posterior $\pi(\kappa \mid \x)$ in terms of the latent variables $\omega_i$.




\section{Topological analysis of image data}

Now we provide a very cursory introduction to the particular TDA method we use in this paper to derive topological information: the persistent homology of images. We then discuss how we go about using the derived persistence diagrams to choose a vectorization (i.e. multivariate feature representation) which captures topological changes in image series.

\subsection{Persistent homology of images}

Often it is convenient to treat a ($k \times l$) image as a vector $\tilde{\bm{x}}_i$ in $\R^{kl}$. However, for the computation of shape information via persistent homology, it is more convenient to treat $\tilde{\bm{x}}_i$ as a function---called the \emph{image map} $I$---with finite, rectangular support on the two-dimensional integer lattice $\mathbb{Z}^2$ which takes values in the extended reals $\bar{\R}$. The sublevel sets of such a function can be considered as binary images, and keeping track of the how the shape information in these images changes through thresholding 
yields the \emph{persistence diagrams} $\mathcal{D}^0$ and $\mathcal{D}^1$, corresponding to 0- and 1-dimensional features respectively. More information on the finer points of persistence diagrams and its underlying theory (persistent homology), appear in Section~S5 of the Supplementary Material. To continue introducing persistence diagrams, we need to define various shape features in a binary image. 

\begin{definition}
In a binary image, a \emph{connected component} (contributing to $\mathcal{D}^0$) is a connected black region and a \emph{loop/hole} (contributing to $\mathcal{D}^1$) is a connected white region surrounded by black pixels. Note that for the purpose of computing homology we consider pixels outside of a binary image as white. 
\end{definition}

In general, a persistence diagram $\mathcal{D}$ is a multiset of points $(b,d)$ in $\R^2$ where the $x$ and $y$ coordinates of $(b,d) \in \mathcal{D}$ correspond to the pixel values at which a shape feature appears (is \emph{born}) and then merges/fills in (\emph{dies})---see Figure S4 for an illustration. To explain this further, we will take $\mathcal{D}^0$ as an example\footnote{As the persistence diagram $\mathcal{D}^0$, and $\mathcal{D}^1$ for the negated image, satisfy a duality property \citep{garin2020duality}, we only describe the persistence diagram $\mathcal{D}^0$.}. For simplicity, we assume that our image map $I$ is injective so that all pixels have a unique intensity value. If the pixel $p$ is such that $I(p)$ is a local minimum, then it will appear at the threshold $b_p$ and not be surrounded by any other black pixels. The pixel $p$ creates or ``gives birth'' to a connected component $C_p$. As the threshold $t$ increases, $C_p$ will continue to gain more and more black pixels until it merges with another connected component $C_q$ at threshold $d$. If the threshold $b_q$ at which $C_q$ appears is less than $b_p$, then we say that $C_p$ ``dies'' at threshold $d_p = d$, and associate the values $(b_p,d_p)$ to the connected component $C_p$. Thus, for the set $M$ of all the local minima in the image we have that 
$$
\mathcal{D}^0= \big\{(b_p, d_p): p\in M\big\},
$$
is the 0-dimensional persistence diagram for $I$. These can be transformed/vectorized in various ways that make them highly useful in statistical/machine learning contexts; one should consult the excellent surveys by \cite{chazal2021} and \cite{top_ml} for further examples beyond those given below.

\subsection{Preliminary image processing} \label{ss:animdat}
Before calculating topological statistics of images in Section~\ref{ss:top_feat}, we must make sure that the image is processed so that the output topological signal is as strong as possible. We consider the simple setup of $n$ image observations $\tilde{\bm{x}}_1, \dots, \tilde{\bm{x}}_n$. As in \cite{detectda}, the images we consider here have been smoothed by a separable Gaussian filter with $\sigma=2$ (which yielded strong results in the same article); however, the $\sigma$ you choose ought to depend on the degree of noise in the image and may be calibrated by the elbow method for a pre-chosen linear combination of topological statistics (ibid.). As we can consider the images as functions from $\mathbb{Z}^2$ to $\bar{\R}$, we may calculate 0 and 1-dimensional sublevel set persistence diagrams according to cubical homology---which we will denote $\mathcal{D}^0_i$ and $\mathcal{D}^1_i$ respectively. Denote $\mathrm{PD}$ to be the space of persistence diagrams. We then calculate some summary $f: \mathrm{PD}^2 \to \R^d$ from this 2-tuple of persistence diagrams to a $d$-tuple of real numbers, $d \geq 1$. Linear functionals of the features 
\[
\bm{x}_i \equiv \psi(\tilde{\bm{x}}_i) := f\big(\mathcal{D}^0_i, \mathcal{D}^1_i\big), \quad i = 1, \dots, n,
\]
are supposed to better represent the change than any linear functionals of the image itself. Though detecting topological change is our main focus, $\psi: \R^{kl} \to \R^d$ can be arbitrary in the exposition below. In particular we assume that if a changepoint $\kappa \in \{1, \dots, n-1\}$ is present then it is represented in the univarate change in distribution of some linear functional of $\psi(\tilde{\bm{x}}_i)$. As the main objective is to recover changes dictated by shape features, throughout this article will we take the mean pixel intensity across the image $\tilde{\bm{x}}_i$ to equal 0, i.e. 
\[
\frac{1}{kl} \sum_{j=1}^{kl} \tilde{\bm{x}}_{ij} \equiv 0,
\]
and the variance of the pixel intensities to equal 1, i.e. 
\[
\frac{1}{kl} \sum_{j=1}^{kl} \tilde{\bm{x}}_{ij}^2 \equiv 1.
\]

\subsection{Crafting a topological feature mapping} \label{ss:top_feat}

For any fixed $d \geq 1$, there are uncountably many distinct functions $f: \mathrm{PD}^2 \to \R^d$ we could choose. Thus, finding a good function $f$ is a nontrivial task. The article by \cite{obayashi2018} served as one of the points of departure for this article. The authors use \emph{persistence images} \citep{pers_image} as their functional which---in concert with logistic regression---yield ``hotspots'' on a dual persistence image reconstructed from $\bm{\beta}$. However, they consider only labelled data for their learning task, and provide no estimates of uncertainty for their recovered coefficients $\bm{\beta}$. We do not use persistence images here, but we do describe a feature embedding $f_{\text{PI}}$ and demonstrate that persistence images are capable of recovering the locations of a topological change in Section S4 of the Supplementary Material. 

The approach that we consider is to select a topological feature embedding (from \citealt{chung2018}) which represents a wide range of topological summary statistics, wherein the statistics we describe were shown to achieve strong test accuracy in a support vector machine classification task of skin lesions. We will use the same suite of persistence statistics used by \cite{chung2018}, along with the ALPS statistic introduced in \cite{detectda}. To further justify our choice in this context, we note that the ALPS statistic and the persistent entropy were demonstrated to capture nanoparticle dynamics well in the videos of \cite{detectda}. Additionally, the same slate of persistence statistics we use here demonstrated superior classification ability in the microscopy imaging study of \cite{pritchard2023}. Furthermore, as the persistence diagrams $\mathcal{D}^0$ and $\mathcal{D}^1$ provide information on the connectivity structure of dark (resp. light) regions in the grayscale image series we analyze, these persistence statistics provide us interpretable information on the distribution of this connectivity.

To detail the persistence statistics we used, first define $l_p = d_p - b_p$ and $m_p = (d_p+b_p)/2$, $p \in M$ and construct our topological embedding $f_{\text{stat}}$ to have the form
\[
f_{\text{stat}}\big(\mathcal{D}^0, \mathcal{D}^1\big) := (T_j)_{1 \leq j \leq 36},
\]
with $T_j$ equal to the following statistics of the empirical distributions of $l_p$ and $m_p$ for the persistence diagrams $\mathcal{D}^0$ and $\mathcal{D}^1$. The various $T_j$ are the means of $l_p$ and $m_p$ for $\mathcal{D}^0$ and $\mathcal{D}^1$; variance of $l_p$ and $m_p$ for $\mathcal{D}^0$ and $\mathcal{D}^1$; skewness of $l_p$ and $m_p$ for $\mathcal{D}^0$ and $\mathcal{D}^1$; kurtosis of $l_p$ and $m_p$ for $\mathcal{D}^0$ and $\mathcal{D}^1$; $25^{th}$, $50^{th}$, and $75^{th}$ percentiles of $l_p$ and $m_p$ for $\mathcal{D}^0$ and $\mathcal{D}^1$; interquartile range of $l_p$ and $m_p$ for $\mathcal{D}^0$ and $\mathcal{D}^1$; persistent entropy of $l_p$ for $\mathcal{D}^0$ and $\mathcal{D}^1$; and ALPS statistic of $l_p$ for $\mathcal{D}^0$ and $\mathcal{D}^1$.

For a large class of persistence statistics, we can establish their stability using Theorem 3 of \cite{divol_polonik} in conjunction with Theorem 5.1 of \cite{skraba2022}. This means that even in the presence of a moderate amount of noise, if a ``separability'' condition holds with high probability (as in Proposition~\ref{p:unimod}), our algorithm will return the correct changepoint as the posterior mode of $\pi(\kappa \mid \bm{\beta}, \x)$.  

\section{Simulation studies} \label{s:sim_app} 

To demonstrate the utility of our method, we consider a simple simulated setup and evaluate our method against other well-known methods in the multivariate changepoint literature. We consider both a straightforward and a more difficult (noisier) changepoint problem for a topological change in a rather short image series. This is to show that our topological feature embedding requires relatively little data to perform well, and that our changepoint method also performs well and is robust in this ``small'' data setting. The following two subsections discuss the ability of our method to detect complicated changes in multivariate data of mixed type as well as changes in covariance. 

\subsection{Detecting a topological change in image series}

For our simulation, we consider a sequence of 50 random images $\tilde{X}_i$, $i = 1, \dots, 50$ of size $50 \times 50$ each consisting of i.i.d. standard Gaussian noise. For ease of notation, we will denote the random image $\tilde{X}_i$ as $X_i$. For ease of exposition we begin by examining our method on a single image series $V^{(1)} = \big(X^{(1)}_i\big)_{1 \leq i \leq 50}$. Initially, we consider a changepoint\footnote{Henceforth, let us denote a \emph{fixed, true} changepoint as $\kappa^*$.} $\kappa^*=25$, whereafter a random rectangular region with intensity of $-2$ is added to each $X^{(1)}_i$. Namely, if $X^{(1)}_{ijk}$ is the pixel at row $j$ and column $k$ of the image $X^{(1)}_i$, we have that 
\[
X^{(1)}_{ijk} = 
\begin{cases}
Z_{ijk} - 2 &\mbox{ if } L_1-W_1 \leq j \leq L_1+W_1, L_2-W_2 \leq k \leq L_2+W_2 \\
Z_{ijk} &\mbox{ otherwise }
\end{cases},
\]
where $Z_{ijk}$ are i.i.d. $N(0,1)$ for $1 \leq i,j,k \leq 50$, $L_m$, $W_m$ are independent and uniformly distributed on $\{5,6, \dots, 43, 44\}$ and $\{2, 3, 4\}$ respectively, for $m = 1,2$. Note that $V^{(1)} = (X^{(1)}_{ijk})_{1 \leq i,j,k \leq 50}$. The additional 1000 videos we simulate according to this formula will be called Experiment 1.

\begin{figure}[t]
\centering
\includegraphics[width=0.8\textwidth]{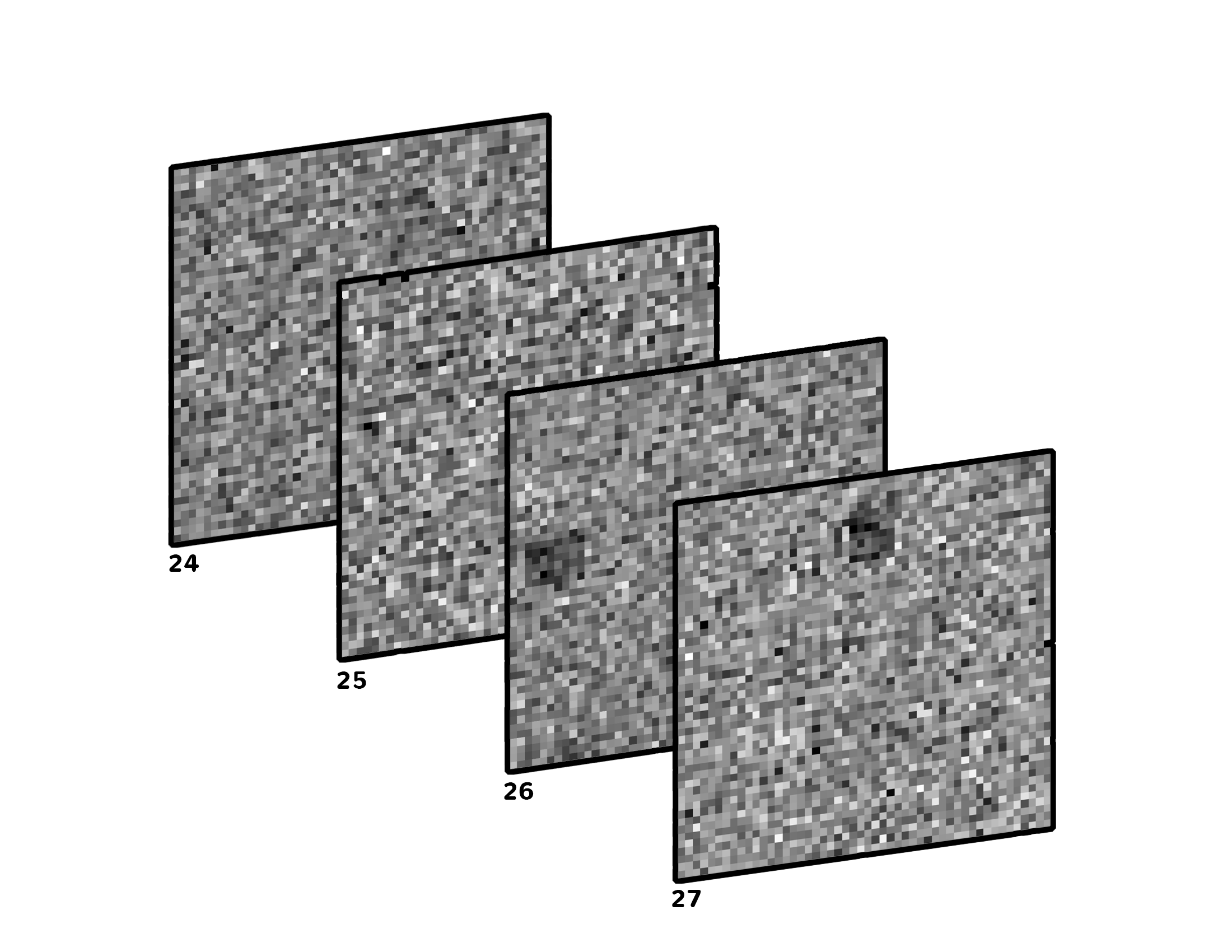}
\caption{Frames 24--27 of the first simulated video $V^{(1)}$ where $\kappa^*=25$ and $n=50$. The changepoint manifests as a darkening of a small region in frames 26 and 27 (and beyond).}
\label{f:before_afterSIM1}
\end{figure}

Examining these images without noise yields a sequence before the changepoint corresponding to no sublevel set homology, and a distribution after the changepoint corresponding to a randomly located connected component with lifetime equal to 2. As mentioned in Section~\ref{ss:animdat}, we standardize each of the images $X_i$ to have mean pixel intensity 0 and standard deviation 1. As such, any estimated change in mean or variance of the image series is entirely spurious and we will be less likely to capture change that is not purely topological. Images of the video before and after the changepoint for this initial scenario can be seen in Figure~\ref{f:before_afterSIM1}.

We compare our method---which we deem BCLR\footnote{Bayesian Changepoint via Logistic Regression.}---to the random forest based classification changepoint detection method from the \emph{changeforest} package \citep{londschien2022} (which we deem CF in the sequel); the E-divisive method (ECP) from the \emph{ecp} package \citep{ecp_jss, matteson2014}; and, the KCP kernel change-point method from \cite{arlot2019} as implemented in the \emph{ruptures} package \citep{selective_review}. To demonstrate the dual benefit of our approach in conjunction with TDA, we gave our method the TDA features from $f_{\text{stat}}$, and the other methods dimensionality-reduced features by first vectorizing the images and then projecting them down onto their first 36 principal components. We did consider a Bayesian changepoint method with available code, but this did not perform well. Information about this and the parameters used for each method can be seen in Section S2 of the Supplementary Material. 

%

We also devised new topologically-aware versions of CF, ECP, and KCP---where we fed these algorithms the same topological features from $f_{\text{stat}}$. We christened these methods CF+TDA, ECP+TDA, and KCP+TDA respectively. For fair comparison, each algorithm received either the exact same PCA or standardized TDA features (depending on which they were designed for). We estimated the changepoint $\hat{\kappa}$ via the posterior mode in our setup, in light of the theory in Section~\ref{ss:theory} and to ensure the estimate corresponded to an actual frame. For experiment 1, we specified our priors to be $N(0, 3I_d)$ for $\bmb$ and discrete uniform on $\{1, \dots, 50\}$ for $\kappa$. For our method, we always took the 2500 posterior samples for both $\kappa$ and $\bmb$ after the burn-in period of 2500 iterations (more details on the properties of the Gibbs sampler and its convergence are available in Section S2.2 of the Supplementary Material).

\begin{figure}
\includegraphics[width=0.49\textwidth]{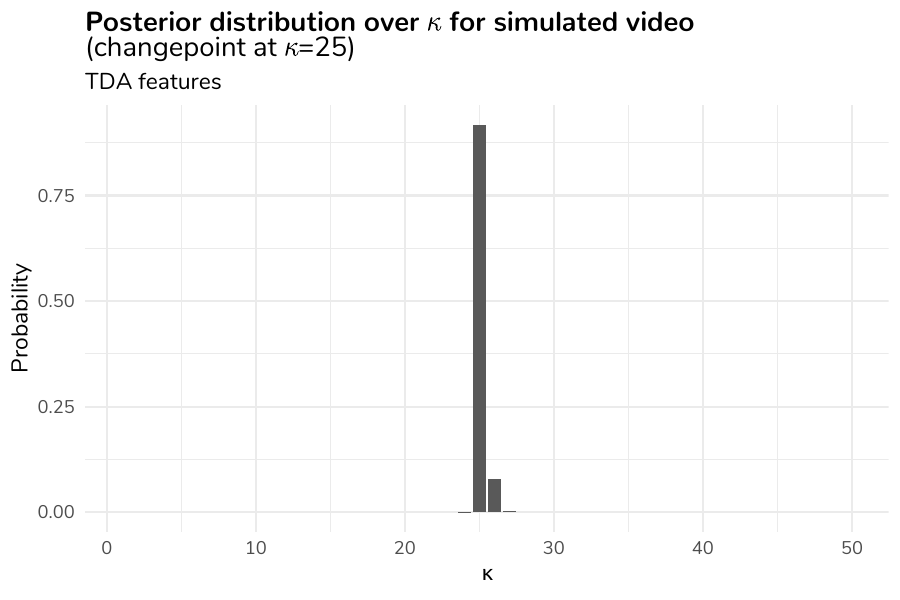}
\hfill
\includegraphics[width=0.49\textwidth]{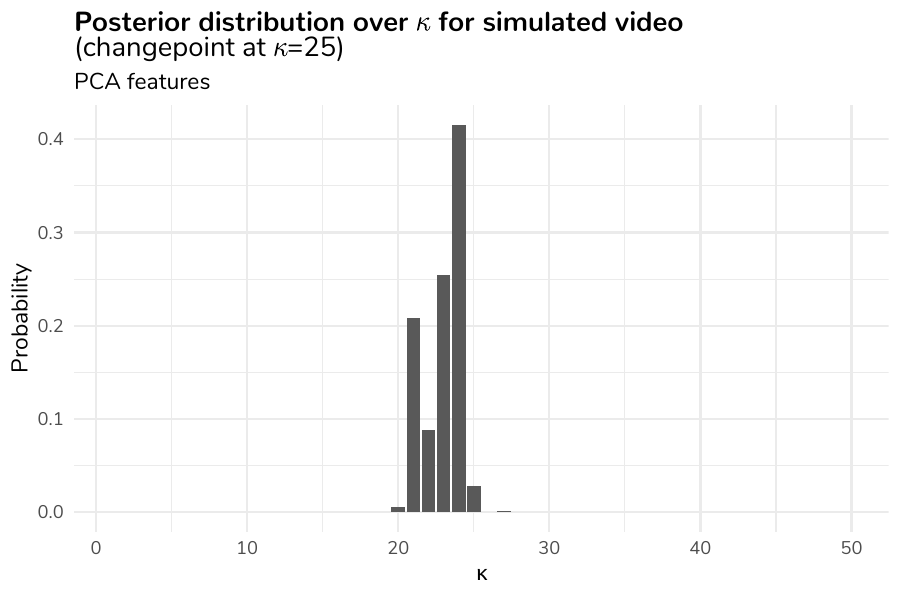}
\caption{The posterior distribution $\pi(\kappa \mid x)$ for the first simulated video $X^{(1)}$ with $\kappa^*=25$. Note that even with the PCA features, our method gets close to the true changepoint at $\kappa^*=25$.}
\label{f:sim_output}
\end{figure}

For the single image series $V^{(1)}$ the estimated posterior change distribution using our algorithm can be seen in 
Figure~\ref{f:sim_output} (left). On the other hand, running CF and ECP algorithms on the PCA features yields estimated changepoints of 1 and 46 for CF and ECP respectively ($p$-value = 0.3 for CF and 0.795 for ECP). There is no $p$-value provided for KCP, but the estimated changepoint was frame 16. However, as one can see from Figure~\ref{f:sim_output} (right), we are not too far from the ground truth if we instead use the PCA features in conjunction with our method. 
The topologically-aware methods CF+TDA, KCP+TDA, and ECP+TDA perform quite well for the image series $V^{(1)}$, recovering the changepoint at $\kappa^*=25$.

We also investigate a second image series $V^{(2)} = (X^{(2)}_{ijk})_{1 \leq i,j,k \leq 50}$ with a changepoint at $\kappa^*=40$ and $X^{(2)}_{ijk} = Z_{ijk} + 1$ instead of $Z_{ijk} - 2$. The additional 1000 videos we simulate according to this formula will be called Experiment 2. For the single random image series $V^{(2)}$ we see similar results those seen in the case of $V^{(1)}$. All of our topological methods detect the changepoint, but at much lower levels of significance across the board ($p$-value = 0.035 for CF+TDA and 0.05 for ECP+TDA). The methods without the TDA features yield estimates of no changepoints for CF and ECP ($p$-values of 0.295 and 0.975, respectively). For $V^{(2)}$, KCP estimates a changepoint of 6. 

To verify that our BCLR method performs well for more than just two specific videos, we generated 1000 additional i.i.d. random image series $V^{(1)}$ (Experiment 1) and $V^{(2)}$ (Experiment 2) and gauged the behavior of the various methods in both experiments. For each of the methods, we calculated the accuracy of the estimated changepoint $\hat{\kappa}$ in terms of proportion of times the estimated changepoint was exactly correct (``\% Exact'') and the root mean-squared error, or ``RMSE'' (of the estimated changepoint from $\kappa^*=25, 40$). For our BCLR method we calculated the RMSE within each of the posterior samples of length 2500 and reported the mean RMSE across all 1000 simulations. Because we calculated the RMSE in this way, we are able to report the standard error of the RMSE. 
In Section S2.4 in the Supplementary Material, we expound on this line of thought and also report the RMSE of the posterior mean and the posterior mode for a few select cases.

We gave the same 36-dimensional standardized TDA features to each of the TDA methods, as well as the same first 36 principal components of the image to the other methods. The results of Experiment 1 can be seen in Table~\ref{t:exp1}.
\begin{table}
\centering
\bgroup
\def\arraystretch{1.25}
\begin{tabular}{l|rrrr|rrr}
Method & BCLR & CF+TDA & ECP+TDA & KCP+TDA & CF & ECP & KCP \\
\hline \hline
\% Exact & 0.697 & 0.714 & 0.663 & 0.673 & 0.020 & 0.036 & 0.009 \\
 & (0.015) & (0.014) & (0.015) & (0.015) & (0.004) & (0.006) & (0.003) \\
 \hline
RMSE & 0.948 & 1.021 & 1.614 & 1.382 & 16.998 & 13.222 & 15.365 \\
& (0.783) & --- & --- & --- & --- & --- & --- \\
\hline \hline
\end{tabular}
\egroup
\caption{\% Exact and RMSE for 1000 simulated videos across five changepoint methods in the case of $\kappa^*=25$ and $X_{ijk} = Z_{ijk} - 2$ for seven different changepoint methods. Standard errors for \% Exact and RMSE indicated in parentheses. SEs for RMSE not applicable for the other methods.}
\label{t:exp1}
\end{table}
First, the methods that we have developed that incorporate the TDA features vastly outperform the ones that only use the PCA features. There hardly seems to be any signal at all for the methods applied to the conventional PCA features, and in fact the results are hardly any different from a uniform random choice. 
Though all of the methods with the TDA features perform well, our method evinces the smallest RMSE. This demonstrates not only the utility of the topological features but the additional ability of our changepoint method to yield consistent results. 

If we denote $p_\ell$ to be the probability mass function of $\pi(\kappa \mid \x_\ell)$ as estimated from the MCMC output, where $\x_\ell$ are the features for simulated video $\ell=1, \dots, 1000$, then we may use $p_\ell$ to derive quantiles 
\[
q_{\alpha}^{(\ell)} := \inf\{\kappa: \sum_{i=1}^{\kappa} p_\ell(i) \geq \alpha\},
\]
and thus form a posterior credible interval 
\[
I_{\alpha, \ell} := \big[q_{\alpha/2}^{(\ell)}, q_{1-\alpha/2}^{(\ell)}\big]
\]
for the true changepoint $\kappa^*=25$ for each simulated video. We examine the coverage probability, i.e. 
\[
\P(\kappa^* \in I_{\alpha}) \approx \frac{1}{1000}\sum_{\ell=1}^{1000} \ind{\kappa^* \in I_{\alpha, \ell}}
\]
for BCLR applied to both the TDA and PCA features in Table~\ref{t:cov_cred}.
\begin{table}
\centering
\bgroup
\def\arraystretch{1.25}
\begin{tabular}{l|rrrrr}
& \multicolumn{5}{c}{$\P(\kappa^* \in I_{\alpha})$} \\
$(1-\alpha)\times100\%$ & 50\% & 80\% & 90\% & 95\% & 99\% \\
\hline \hline
TDA & 0.852 & 0.933 & 0.964 & 0.972 & 0.994 \\
PCA & 0.312 & 0.509 & 0.594 & 0.645 & 0.721 \\
\end{tabular}
\egroup
\caption{Estimated credible interval coverage probabilities for $\kappa^*=25$ for our BCLR method. }
\label{t:cov_cred}
\end{table}
Even though $\kappa$ is discrete, the credible intervals for the TDA features in our setup are conservative at each setting of $\alpha$ that we consider. Though the specified intervals do not necessarily have the highest posterior mass (we do this in Table S1 and see similar results), Table~\ref{t:cov_cred} indicates that interval estimation using BCLR with TDA features is appropriate here (where the signal is fairly strong relative to the noise), in a way that imposes no distributional restrictions on the data $\x$. A more detailed analysis of these intervals can be seen in Section S3 of the Supplementary Material. For Experiment 2 ($\kappa^*=40$) the signal is cut in half, and corresponding the coverage probabilities\footnote{The study of frequentist coverage probability bias of these types of Bayesian posterior quantile intervals was studied in \cite{sweeting2001}, albeit those results are not applicable to this more complicated setting.} are less than their nominal amount---see Table~S2. Nevertheless, the probability our method will return an estimate containing the true changepoint is higher than the other methods due to our ability to seamlessly carry out interval estimation. 

As a proof-of-concept, we may use the posterior sample estimates of the mean and covariance we gained from the single image series $V^{(2)}$ as a prior for $\bmb$ for the 1000 videos in Experiment 2. We considered this additional prior for $\bmb$ for our method to see if this additional information provides us further ability to discriminate the changepoint location accurately. What we call the ``data-driven prior'' was a multivariate normal with mean equal to the posterior sample mean and covariance equal to the sample covariance of $\bmb$ from the initial simulated video with $\kappa^*=40$. Along with this prior for $\bmb$ we considered a ``binomial'' prior for $\kappa$, where 
$$
\pi(\kappa) = \dbinom{48}{\kappa-1}(0.8)^{\kappa-1}(0.2)^{48-\kappa+1}, \quad \kappa = 1, \dots, 49
$$
so that the unique mode of $q$ is $\kappa = 40$ and no prior mass is placed on $\kappa=50$. For this experiment, we also conducted 5000 Monte Carlo iterations and chose the 2500 simulations after the burn-in. Additional details on the performance of this method with alternative priors for $\bmb$ and $\kappa$ can be seen in Section S2.1 of the Supplementary Material. 
As one can see in Table~\ref{t:exp2}, our method using the priors specified above outperforms both KCP+TDA and ECP+TDA methods and vastly outperforms the non-topological versions of those methods as well as all versions of CF.

\begin{table}
\centering
\bgroup
\def\arraystretch{1.25}
\begin{tabular}{l|rrrr|rrr}
Method & BCLR-* & CF+TDA & ECP+TDA & KCP+TDA & CF & ECP & KCP  \\
\hline \hline
\% Exact & 0.583 & 0.458 & 0.520 & 0.534 & 0.020 & 0.020 & 0.019  \\
 & (0.016) & (0.016) & (0.016) & (0.016) & (0.004) & (0.004) & (0.004) \\
 \hline 
RMSE & 1.537 & 5.030 & 4.330 & 3.300 & 21.191 & 19.009 & 21.729 \\
& (1.032) & --- & --- & --- & --- & --- & --- \\
\hline \hline
\end{tabular}
\egroup
\caption{\% Exact and RMSE for 1000 simulated videos across six changepoint methods in the case of $\kappa^*=40$ and $X_{ijk} = Z_{ijk} + 1$. The method BCLR-* corresponds to our method with a data-driven prior for $\bmb$ and a binomial prior for $\kappa$ with mode at $\kappa=40$. Standard errors for \% Exact and RMSE indicated in parentheses. SEs for RMSE not applicable for the other methods.}
\label{t:exp2}
\end{table}



Furthermore, none of other methods we describe tell us what coordinates are most important for detecting such a change. With regard to the 1000 simulated videos in Experiment 1 ($\kappa^*=25$) we measured the importance of a given coordinate of $\bmb$, with respect to the TDA features, via the signal-to-noise ratio (SNR)\footnote{This quantity is meaningful as each coordinate has mean 0 and variance 1. It is defined as $\bar{\beta_j}^2/s(\beta_j)^2$, $j = 1, \dots, d$.}. We can see the mean and standard deviation of the SNRs---for persistence statistics which had the highest SNR most often among the 1000 simulations---in Table~\ref{t:sim1000}. We were also able to get the mean posterior correlations for the regression coefficients corresponding to the persistence statistics. The largest absolute mean correlation among the 5 statistics in Table~\ref{t:sim1000} was -0.358 between kurtosis and skewness of the lifetimes for $\mathcal{D}_0$. The smallest absolute mean correlation was -0.007 between the ALPS statistic and the persistent entropy. The other absolute mean correlations hovered between 0.043 and 0.2, indicating that there was not necessarily a single statistic that stood out above the rest and justified the use of the feature embedding $f_{\text{stat}}$.

To demonstrate the effectiveness of the proposed method beyond the context of image series, we conduct additional simulation studies focusing on two other important changepoint settings. The first simulation study shows that BCLR can reliably detect and characterize a change in distribution for data with both continuous and discrete components. The second simulation study illustrates how, after an appropriate feature mapping, BCLR can determine the location and precise nature of a change in covariance. 


\begin{table}
\centering
\bgroup
\def\arraystretch{1.25}
\begin{tabular}{l|r|r}
Statistic & Prop. highest SNR & Mean (SD) Posterior SNR  \\
\hline
\textbf{persistent entropy} of $l_p$ for $\mathcal{D}^0$ & 0.185 & 1.562 (1.082)  \\
\textbf{skewness} of $l_p$ for $\mathcal{D}^0$ & 0.156 & 1.873 (0.817)  \\
\textbf{kurtosis} of $l_p$ for $\mathcal{D}^0$ & 0.120 & 1.734 (0.875) \\
\textbf{ALPS statistic} of $l_p$ for $\mathcal{D}^0$ & 0.114 & 1.118 (1.218) \\
\textbf{variance} of $l_p$ for $\mathcal{D}^0$ & 0.111 & 1.703 (0.683) \\
\end{tabular}
\caption{Table of posterior coefficient signal-to-noise ratios for the 1000 simulated videos in Experiment 1 along with the proportion (out of 1000 simulations) in which said statistic had the highest SNR.}
\label{t:sim1000}
\egroup
\end{table}







\subsection{Detecting a change in data of mixed type}\label{ss:mixed}

To evaluate the performance of BCLR in detecting and characterizing a change in data of mixed type, we simulated 2500 independent sequences of the form $X_{1}, \dots, X_{600}$ where each observation $X_i = (X_{i1}, X_{i2}, X_{i3}, X_{i4}, X_{i5})$, $i = 1, \dots, 600$ is five-dimensional. The coordinates of $X_i$ are defined by $(X_{i3}, X_{i4}, X_{i5}) = (Y_{i3}, Y_{i4}, Y_{i5})B$
where each of the $Y$-coordinates are i.i.d. standard Laplace random variables and $B$ is the matrix defined by
\[
B:=
\begin{pmatrix}
1 & 0 & 0 \\
0 & 2 & -1 \\
0 & 0 & 1 
\end{pmatrix}.
\]
We chose the Laplace distribution to see how our method would fare in the presence of heavier-tailed noise and we chose $B$ to induce dependence between coordinates 4 and 5, to make the changepoint task more difficult. Coordinates $X_{i1}$ and $X_{i2}$ correspond to a dummy coding of categorical variables $Z_i \in \{1,2,3\}$ with distribution $\P_0$ until frame 350 and distribution $\P_1$ after frame 350---where 
\[
\P_0(Z_i=1) = 0.5 \text{ and } \P_0(Z_i=2) = 0.2,
\]
and 
\[
\P_1(Z_i=1) = 0.1 \text{ and } \P_1(Z_i=2) = 0.5,
\]
and $X_{i\nu} = \ind{Z_i = \nu}$, for $\nu = 1,2$ and all $i$. 

\begin{figure}[t]
    \centering
    \includegraphics[width=0.6\textwidth]{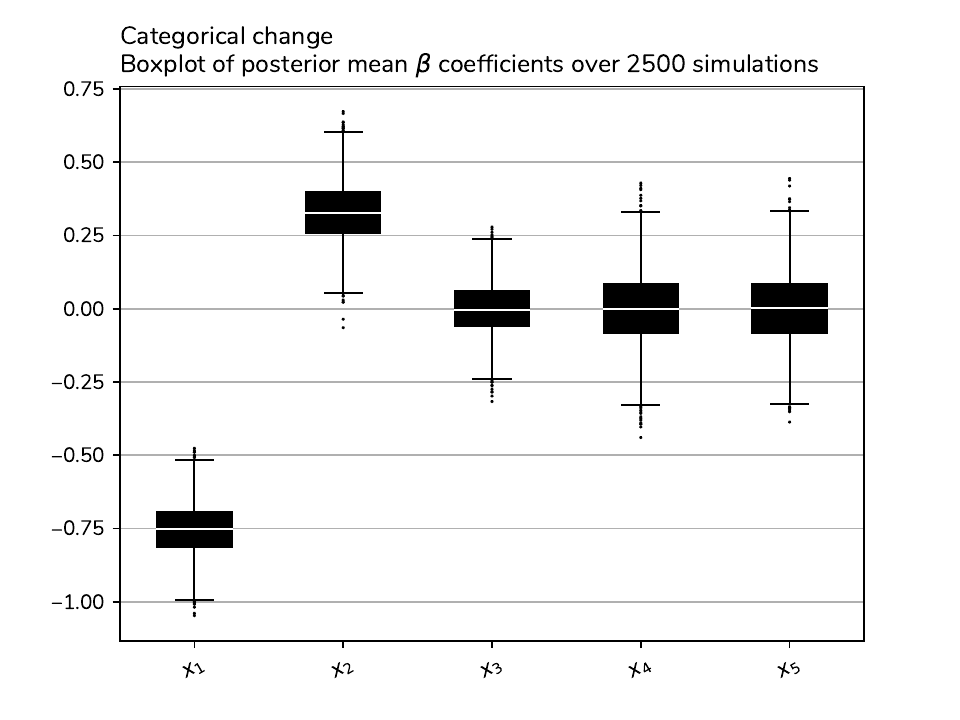}
    \caption{Distribution of the posterior means of $\bmb$ (on standardized scale) over all 2500 simulations for the mixed data change setup of Section \ref{ss:mixed}. This figure clearly indicates a strong influence of the component of $\bmb$ associated with a drop in the value of $P(Z=1)$ after the change (via $x_1$).} 
    \label{f:post_mean_mixed}
\end{figure}

We compare our BCLR method to five alternatives. Three of the methods we have already encountered: CF, ECP, and KCP. We chose a uniform prior for $\kappa$ and a $N(0, 3^{-1}I_d)$ prior for $\bmb$. This added regularization was to ensure less variance in the posterior estimates of $\bmb$ and hence better results in practice. The MEAN changepoint method mentioned in Table~\ref{t:exp4} finds the best single change which minimizes the $L^2$ cost function---see Chapter 3 of \cite{chen2012}. Here we used the default settings for each changepoint method (which we detail in Section S2) and run the BCLR method for our standard 5000 iterations, with 2500 post burn-in taken as our posterior sample. As one can see in Figure~\ref{f:post_mean_mixed}, the BCLR method detects that the probability for $Z=1$ decreases and that the probability for $Z=2$ increases after the changepoint. It worth reiterating that our method ascertains the nature of this change without any labels which indicate whether or not an observation is before or after the changepoint. Here, our method performs comparably to ECP and KCP, beats out CF, and vastly outperforms methods which are given only the raw (non-standardized) data.

\begin{table}[t]
\centering
\bgroup
\def\arraystretch{1.25}
\begin{tabular}{l|rrrrrrr}
Method & BCLR & CF-raw & ECP-raw & ECP & KCP-raw & KCP & MEAN-raw \\
\hline \hline
\% Exact & 0.260 & 0.216 & 0.003 & 0.275 & 0.102 & 0.266 & 0.020 \\
& (0.009) & (0.008) & (0.001) & (0.009) & (0.006) & (0.009) & (0.003) \\
\hline
RMSE & 6.796 & 11.306 & 167.851 & 6.375 & 54.153 & 6.581 & 182.753 \\
& (5.437) & --- & --- & --- & --- & --- & --- \\
\hline \hline
\end{tabular}
\egroup
\caption{Comparison of various changepoint detection methods in the discrete variable change setup. Here ``raw'' indicates the method was fed data that was NOT centered to have mean zero and standardized to have standard deviation 1 in each coordinate. Standard errors for \% Exact and RMSE indicated in parentheses. SEs for RMSE not applicable for the other methods.}
\label{t:exp4}
\end{table}

\subsection{Detecting a change in covariance} \label{ss:cov_ch}

\begin{figure}
    \centering
    \includegraphics[width=0.6\textwidth]{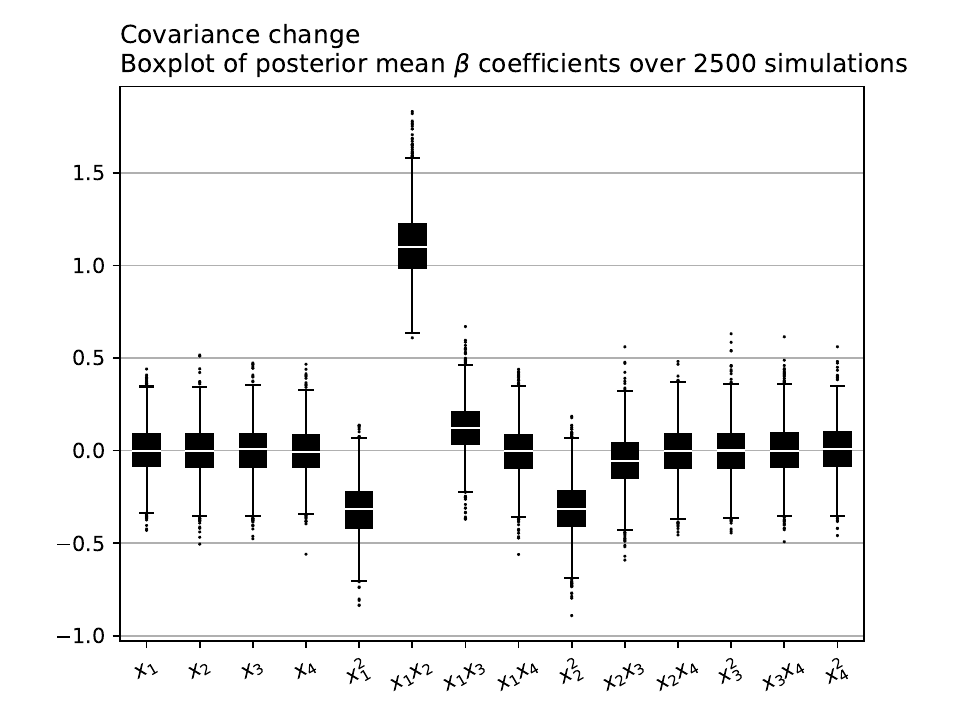}
    \caption{Distribution of the posterior means of $\bmb$ (on standardized scale) over all 2500 simulations for the covariance change setup of Section \ref{ss:cov_ch}. This figure clearly indicates a strong influence of the component of $\bmb$ associated with the product term $x_1x_2$, which correctly suggests an increase in the correlation between the first two coordinates of $X_i$.}
    \label{f:post_mean_cov}
\end{figure}

To conclude our experiments in the single changepoint setup, we study the rather difficult problem of detecting a change in covariance. As in the previous subsection, we simulated 2500 independent sequences $X_{1}, \dots, X_{300}$
where $X_{i} \overset{\text{ind}}{\sim} N(0, \Sigma_i)$ for all $i = 1, \dots, 300$. We define the covariance matrix
\[
\Sigma_i =
\begin{cases}
\Sigma_0 &\mbox{ if } j \leq 200 \\
\Sigma_1 &\mbox{ if } j > 200,
\end{cases}
\]
with $\Sigma_0 = I_4$ (i.e. $4\times 4$ identity matrix) and 
\[
\Sigma_1
=
\begin{pmatrix}
1 & 0.8 & 0.1 & 0 \\
0.8 & 1 & 0 & 0 \\
0.1 & 0 & 1 & 0 \\
0 & 0 & 0 & 1
\end{pmatrix}.
\]
We consider the following degree-2 polynomial feature embedding of $x = (x_1, x_2, x_3, x_4)$
\[
\psi(x) := (x_1, x_2, x_3, x_4, x_1^2, x_1x_2, x_1x_3, x_1x_4, x_2^2, x_2x_3, x_2x_4, x_3^2, x_3x_4, x_4^2) 
\]
so that $\psi: \R^4 \to \R^{14}$. Per usual, we standardize\footnote{That is, subtract from each column the mean of the series and then divide by the series' standard deviation.} each series $( \psi(X_{i}), i = 1, \dots, 300 )$ prior to feeding it to our algorithm. For our method, we used the same priors as Section~\ref{ss:mixed} and ran the algorithm for 5000 iterations, discarding the first 2500 as burn-in. We compare our BCLR method to five alternatives and use the same parameters for these methods as the previous section. We apply all three of these methods to the standardized data. We also applied CF and a gaussian likelihood-based method (\citealp{lavielle2006}, which we deem GAUSSIAN) to the raw data $(X_{i}, i = 1, \dots, 300)$ as a basis of comparison. The results of the experiment for all methods are seen in Table~\ref{t:exp3}, and the boxplot of the posterior means of the $\bmb$ coefficients from our method are seen in Figure~\ref{f:post_mean_cov}. We see that our method greatly outperforms all but the GAUSSIAN changepoint method, which ought to work well as it satisfies the parametric assumptions of the problem.

\begin{table}[t]
\centering
\bgroup
\def\arraystretch{1.25}
\begin{tabular}{l|rrrrrr}
Method & BCLR & CF & ECP & KCP & CF-raw & GAUSSIAN-raw \\
\hline \hline
\% Exact & 0.150 & 0.115 & 0.018 & 0.063 & 0.066 & 0.300 \\
& (0.007) & (0.006) & (0.003) & (0.005) & (0.005) & (0.009) \\
\hline
RMSE & 14.710 & 35.042 & 77.073 & 72.969 & 63.925 & 4.756 \\
& (12.696) & --- & --- & --- & --- & --- \\
\hline \hline
\end{tabular}
\egroup
\caption{Comparison of various changepoint detection methods in the covariance change setup. The BCLR method performs better than all but the correctly-specified Gaussian covariance method. Here ``raw'' indicates the method was fed the raw data, rather than the standardized degree-2 polynomial features. Standard errors for \% Exact and RMSE indicated in parentheses. SEs for RMSE not applicable for the other methods.}
\label{t:exp3}
\end{table}

\section{Applications}

In this section we discuss the utility of our method to detect topological changes in image series on real world data. The first subsection discusses the ability of BCLR to detect a change in a nanoparticle video and the second subsection demonstrates the ability of the algorithm to detect a solar flare event, hinting at doing so before the image intensity peaks. In the Supplementary Material, we apply our multiple changepoint extension to the full (1659--2023) Central England Temperature series dataset \citep{cet}.

\subsection{Structural change in a nanoparticle video}

\begin{figure}[t]
\centering
\includegraphics[width=0.98\textwidth]{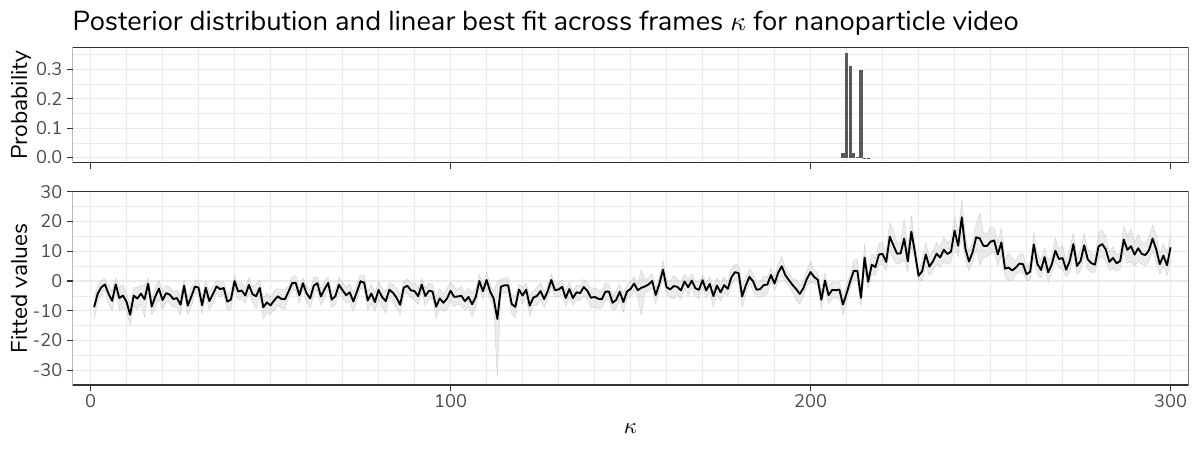}
\caption{Nanoparticle data. (Top) the posterior distribution (over last 2500 Gibbs sampler draws) of $\kappa$. (Bottom) the fitted values $\bm{x}_{\kappa}^\top\bm{\beta}$ for the last 2500 draws of $\bm{\beta}$. 95\% credible bands depicted in light gray.}
\label{f:nano_post}
\end{figure}

\begin{figure}[t]
\centering
\includegraphics[width=0.98\textwidth]{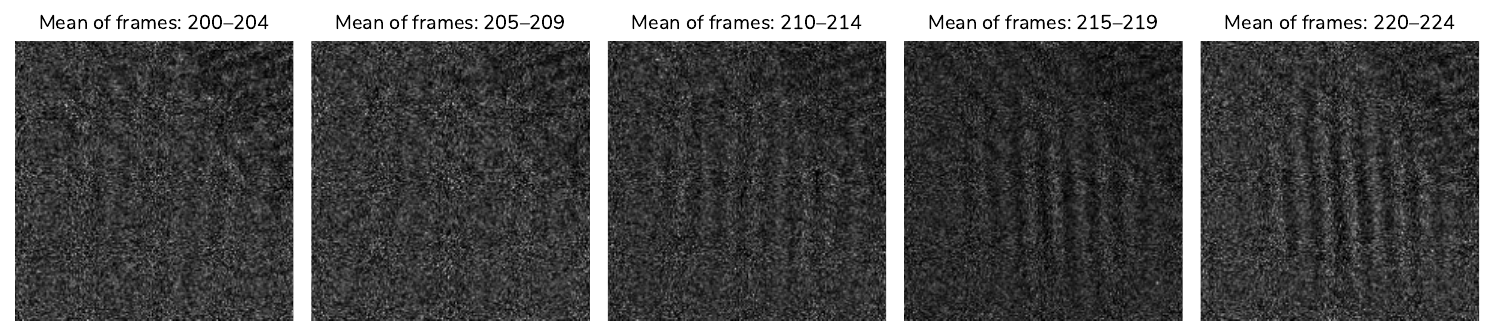}
\caption{Frame-averaged nanoparticle video. One can see the lack of crystalline structure prior to frame 210 and the presence of structure afterwards.}
\label{f:nano_mean}
\end{figure}

Evidence presented in \cite{detectda} indicates that a reasonable topological summary for the detection of nanoparticle dynamics is some linear combination of the ALPS statistic (ibid.) and persistent entropy \citep{stab_pers_entr}. We assess the accuracy of this statement using the topological embedding $f_{\text{stat}}$. Visual inspection of the nanoparticle video of interest\footnote{Described in detail in the Supplementary Material of \cite{detectda}.} suggests that a change occurs in the vicinity of frame 210, though the low signal-to-noise ratio precludes any notion of ``ground truth''. Thus, we apply our algorithm to the data to get an estimate of where the change occurs and what the best representation for said change is. Plots describing these results can be seen in Figure~\ref{f:nano_post} and averages of 5 consecutive frames (to improve visualization) before and after the estimated changepoint can be seen in Figure~\ref{f:nano_mean}. We chose a prior for $\bmb$ with mean 0 and covariance matrix equal to $3 I_d$, where $I_d$ is the $d$-dimensional identity matrix. 

As seen in Figure~\ref{f:nano_post}, the marginal posterior distribution of $\kappa$ concentrates around frame 210. After having run our Gibbs sampler for 5000 iterations and discarding the first 2500 samples from the posterior Monte Carlo sample, we estimate that $\pi(210 \mid \x) = 0.3560$, $\pi(211 \mid \x) = 0.3112$, and $\pi(214 \mid \x) = 0.2984$ with probability less than $0.02$ elsewhere. We can see using the posterior mean of $\bmb$ that reasonable separation is achieved for $\bm{x}_{\kappa}^{\top}\bmb$ in Figure~\ref{f:nano_post} (bottom). The question remains as to which topological statistics best represent the change. We summarize the importance of a given coordinate of $\bmb$ by its signal-to-noise ratio (SNR), defined in the previous section. The top 5 statistics in terms of signal-to-noise ratio can be seen in Table~\ref{t:nano}.

\begin{table}[t]
\centering
\bgroup
\def\arraystretch{1.25}
\begin{tabular}{l|r|r}
Statistic & Posterior SNR & Posterior mean \\
\hline
$\bm{25^{th}}$ percentile of $l_p$ for $\mathcal{D}^1$ & 5.694 & 1.087\\
\textbf{variance} of $l_p$ for $\mathcal{D}^0$ & 3.772 & 2.389 \\
\textbf{persistent entropy} of $l_p$ for $\mathcal{D}^1$ & 3.407 & -1.481 \\
\textbf{persistent entropy} of $l_p$ for $\mathcal{D}^0$ & 2.117 & -0.905 \\
\textbf{ALPS statistic} of $l_p$ for $\mathcal{D}^0$ & 1.890 & -1.222 \\
\end{tabular}
\egroup
\caption{Table of posterior coefficient statistics for nanoparticle video.}
\label{t:nano}
\end{table}

The results of Table~\ref{t:nano} support---but also refine---the results of \cite{detectda}, wherein the persistent entropy and the ALPS statistic of the lifetimes of $\mathcal{D}^0$ were chosen to represent the dynamics of the nanoparticle. These results seem to suggest that the variance of $l_p$ of $\mathcal{D}^0$ captures the dynamics even better, having a large absolute posterior mean and rather low variance. For the sake of comparison, we ran ECP+TDA, CF+TDA, and KCP+TDA and recovered an estimated changepoint of $\hat{\kappa} = 214$ for all three methods with $p$-values of 0.005 for the first two. 

\begin{figure}[t]
\centering
\includegraphics[width=0.98\textwidth]{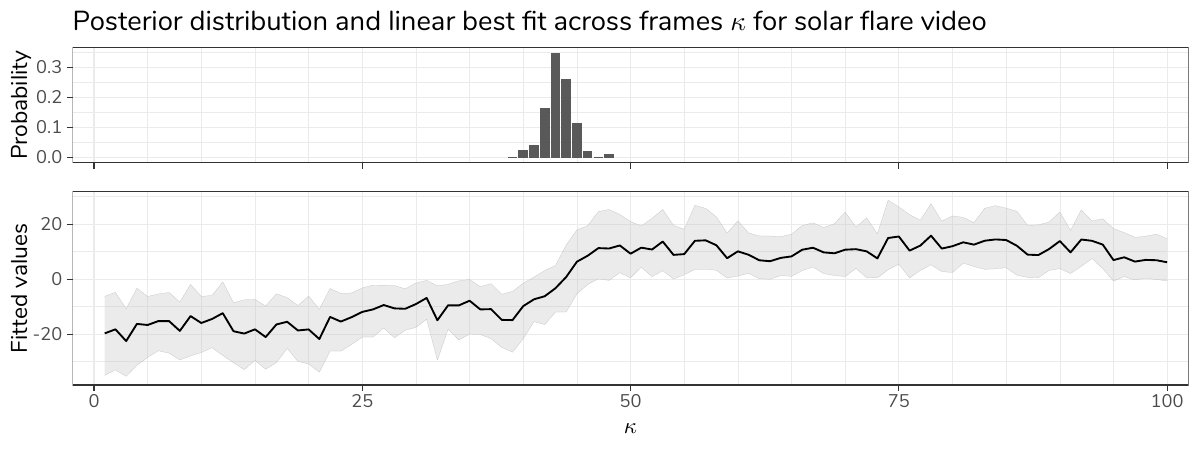}
\caption{Solar flare data. (Top) the posterior distribution (over last 2500 Gibbs sampler draws) of $\kappa$. (Bottom) the fitted values $\bm{x}_{\kappa}^\top\bm{\beta}$ for the last 2500 draws of $\bm{\beta}$. 95\% credible bands depicted in light gray.}
\label{f:solar_post}
\end{figure}

\begin{figure}[t]
\centering
\includegraphics[width=0.98\textwidth]{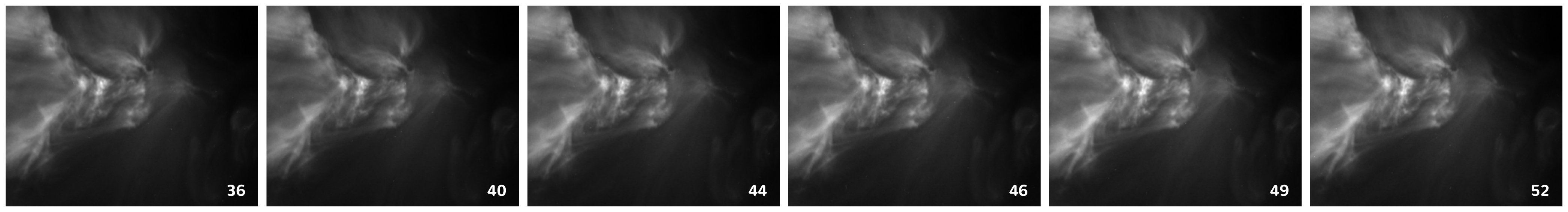}
\caption{Standardized solar flare images. Flare event seems to occur in early portion of 40s frames. The change in intensity seems to be pre-empted by a topological change, as detected by BCLR.}
\label{f:solar_vid}
\end{figure}

\subsection{Solar flare detection}

We conclude this section by looking at a 100 frame solar flare video, which was taken from \cite{xie2012} and analyzed via the topological online changepoint method PERCEPT proposed in \cite{percept}. Frames that depict the video immediately prior to and after the changepoint can be seen in Figure~\ref{f:solar_vid}. Running our changepoint method with the same prior as in the case of the nanoparticle video, we observe a posterior mode of $\hat{\kappa}=44$ using BCLR with $f_{\text{stat}}$ (see histogram (top) in Figure~\ref{f:solar_post}). Note that the change that occurs in Figure~\ref{f:solar_vid} appears to be a sort of gradual change from frames 40 to 49, and this is captured in the posterior distribution in the top figure of Figure~\ref{f:solar_post}. The support of said posterior ranges from around frame 40 to 49, peaking at frame 44, which is right around the midpoint of this transition. One can constrast this with the more highly peaked posterior in the top part of Figure~\ref{f:nano_post}---reflecting the more abrupt change in the video in that application.

To confirm these results, we can apply BCLR to other series that may capture the solar flare event, such as the pixel means of the images. Applying BCLR to this one-dimensional signal\footnote{Here we do not standardize each frame to have mean zero intensity.} returns an estimated posterior mode of $\pi(\kappa \mid \x)$ of $\kappa = 46$. This seems reasonable given the below results for the other methods. In the article \citep{percept}, the changepoint was estimated to be $\kappa = 49$ using PERCEPT. Since they were using a CUSUM approach on their derived topological features, their estimate was more likely to occur after the best separation between pre- and post-change distributions. Also, there is no indication in the article that their images were standardized, which leads one to believe that their estimated changepoint captures mostly the increase in the mean intensity of the solar flare video. That being said, using the PCA features described below, our method has posterior mode $\hat{\kappa} = 49$. The discrepancy between the estimate of the changepoint using topological statistics and pixel means suggests that there is topological change occurring prior to the solar event which ultimately takes place and leads to a great increase in radiation which is visually apparent. 

We also extract 30 principal components from the solar flare image data, as in \cite{percept}. This yields an estimated 17 changepoints at significance level $\alpha=0.02$ using CF. However, the first changepoint is estimated to be 47. The PCA features yield estimated changepoints at $\kappa =33$ and 63 for ECP (with $p$-values 0.005 for each). The method KCP yields an initial changepoint of 45 using the PCA features. Using these PCA features with BCLR tells us there is a posterior probability of at least 95\% that the changepoint lies in the interval $[46, 54]$ (using the posterior quantile intervals above). This robustness---in terms of increased uncertainty---to misspecification of feature embedding can be very useful when we are not dealing with image data, are worried about false positives, or surmise there is no topological change.

\section{Extension of BCLR to multiple changepoints}

Here we discuss the details of our multiple changepoint version of BCLR. We devised a 
versatile extension which retains the 
benefits of the single changepoint setup. We omit the performance experiments in this section for brevity. However, our method is shown to either match or outperform the others in terms of the Rand index \citep{rand1971} and the adjusted Rand index \citep{adj_rand} on a changepoint problem that includes a change in variance, mean, and covariance in a series of only 250 observations. We also apply our method to the full (1659--2023) univariate Central England Temperature series dataset \citep{cet} to demonstrate the ability of our method to work on data with autocorrelation or a trend. We show that the method produces reasonable off-the-shelf results and can provide post-hoc uncertainty quantification for other methods \citep{shi2022}. The results of both experiments can be found in Section S6 of the Supplementary Material.

A practical way to extend BCLR to the multiple changepoint setting is to find a reasonable partition of the indices of the data and apply the single changepoint method 
to each pair of consecutive ``blocks''. Though it would be desirable to have a fully quasi/generalized Bayesian method utilizing multinomial logistic regression, this approach presents its own unique challenges and warrants a more extensive treatment than is possible in this article. 
Given that the Gibbs sampler in the multinomial logistic regression setup would iteratively search for changes in consecutive blocks (which we deem ``segments''), our method can be seen as a rough approximation of that more elegant formulation---see Section S6.1 in the Supplementary Material for more details.

Our multiple changepoint version of BCLR proceeds as follows. First fix some partition of $\{1, \dots, n\}$ (we describe how to do so in the next subsection). Namely, for some $J > 0,$ choose blocks
$$
\{\tau_0+1, \dots, \tau_1\}, \{\tau_1+1, \dots, \tau_2\}, \dots, \{\tau_{J}+1, \dots, \tau_{J+1}\}
$$
where $0 \equiv \tau_0 \leq \tau_1 < \tau_2 < \cdots < \tau_J < \tau_{J+1} \equiv n$. Denote a \emph{segment} as the union of two consecutive blocks, e.g. $\{\tau_{2}+1, \dots, \tau_{4}\}$. From here, we apply BCLR to the datasets (i.e. submatrices of $\bm{X}$ consisting of rows associated with a segment) 
$$
\bm{X}_{j} = 
\begin{pmatrix} 
\bm{x}_{\tau_{j-1}+1}^{\top} \\
\vdots \\
\bm{x}_{\tau_{j+1}}^{\top}
\end{pmatrix}
$$
for $j = 1, \dots, J$. Again, this approximates the procedure that one would use in the multinomial logistic regression setup. From this, one gets $J$ posterior distributions $\pi_j(\kappa, \bmb | \bm{X}_j)$ having support $\{\tau_{j-1}+1, \dots,  \tau_{j+1}\} \times \R^d$. We will denote the probability measure associated with the marginal posterior $\pi_j(\kappa | \bm{X}_j)$ as $\mu^{\kappa}_j$.

There are a few additional matters to address. Denote the estimated changepoint on the $j^{\text{th}}$ segment $\{\tau_{j-1}+1, \dots, \tau_{j+1}\}$ to be $\hat{\kappa}_j$. Based on the construction thus far, it is possible that $\hat{\kappa}_{j} \leq \hat{\kappa}_{j-1}$, especially if we use the posterior mode for estimating the $j^{\text{th}}$ changepoint $\kappa_j$. Suppose that we set some minimum distance $\Delta \geq 1$ between estimated changepoints; i.e. we require that
$$
\hat{\kappa}_{j} - \hat{\kappa}_{j-1} \geq \Delta,
$$
for $j = 1, \dots, J$. Then, in the case where the estimated changepoints correspond to the posterior modes, we can fix $\hat{\kappa}_0 = \tau_0 = 0$ and for $j = 1, \dots, J$ set
\begin{equation}\label{e:kappa_delta}
\hat{\kappa}_{j} = \argmax_{\substack{\kappa \geq \hat{\kappa}_{j-1}+\Delta, \\ \kappa \leq \tau_{j+1} - \Delta}} \pi_{j}(\kappa \mid \bm{X}_j).
\end{equation}
In the case that the interval $[\hat{\kappa}_{j-1} + \Delta, \tau_{j+1}-\Delta]$ is empty we will set $\hat{\kappa}_j \equiv \hat{\kappa}_{j-1}$. Our final estimate of changepoints is then 
$$
\hat{K} := \{\hat{\kappa}_1, \dots, \hat{\kappa}_J\},
$$
so it is possible that $|\hat{K}| < J$. We will discuss how to handle this in the ensuing sections. With this in mind, it is paramount to find a reasonable partition $\tau_1, \dots, \tau_J$---which we will now consider.

\subsection{The warm-up period}\label{ss:warm_up}

We can find a reasonable partition $\tau_1 < \cdots < \tau_J$ by running our algorithm on the datasets $\bm{X}_j$, $j = 1,\dots ,J$ with initial values of segment boundaries $\tau_i = \lfloor nj/(J+1) \rfloor$ and $j = 0, 1, \dots, J+1$. To ensure that each $[\tau_{j-1} + \Delta, \tau_{j+1}-\Delta]$ is non-empty, we will make sure to set $\Delta < \lfloor n/(2J+2) \rfloor$ so that $\Delta \leq \lfloor n/(J+1) \rfloor - \Delta$. 
We will run BCLR on each of the above datasets $\bm{X}_j$, $j = 1, \dots, J$ for a small number of iterations 
to find reasonable values of $\tau_1, \dots, \tau_{J}$ based on the $\hat{K}$ returned by the algorithm during this initial ``warm-up period''. In our experience, the Gibbs sampler converges rapidly. As above, we will remove from consideration any estimated changepoints that do not satisfy the minimum distance requirements of \eqref{e:kappa_delta}. As such, it is worthwhile to set $J > J^*$, where $J^*$ is the true number of changepoints, so that the partition returned by this warm-up period consists of at least $J^*$ segments. As our method works best when there is a most a single changepoint in a segment $\{\tau_{j-1}+1, \dots, \tau_{j+1}\}$, we would like to have $J$ large so that multiple changepoints do not appear in a single dataset $\bm{X}_j.$ 

With the exception of choosing a fixed mesh size of approximately $n/(J+1),$ our approach is similar to the seeded binary segmentation of \cite{kovacs2023}, which searches a collection of fixed intervals of varying resolutions for changepoints and eliminates the changepoints which have the lowest likelihood or ``gain''. We could implement this for BCLR as well, but we retain the idea of searching segments at a fixed resolution owing to parallels with the multinomial logistic setup (cf. Section S6.1 of the Supplementary Material).

\subsection{Bottom-up segmentation}

Having estimated changepoints $\hat{K}$ and calculated marginal posterior distributions $\pi_j(\kappa | \bm{X}_j)$ for each, we would now like to eliminate those $\kappa_j$ with less concentrated posterior distributions. As mentioned in Section S1.1 of the Supplementary Material, BCLR will produce a discrete uniform distribution for a constant signal. On the other hand, if the posterior distribution concentrates at a single value, there is complete certainty as to the changepoint's location. Consider the set $A = \{a+1, a+2, \dots, a+m\}$ for an integer $a$ and positive integer $m$. The discrete uniform distribution $\mu_{\text{unif}} = \frac{1}{b-a}\sum_{i=1}^m \delta_{a+i}$ on $A$ and a point mass $\delta_{b}$ (with $b \in A$) have the maximum and minimum Shannon entropy of any discrete distribution on $A$, respectively. As the distance between estimated changepoints may be unequal, we normalize the entropy of a probability measure $\mu = \sum_{i=1}^m p_i\delta_{a+i}$ on $A$ and denote this normalized entropy as 
$$
\tilde{H}(\mu) := - \frac{1}{\log m}\sum_{i=1}^m p_i \log(p_i),
$$
so that $\tilde{H}(\mu_{\text{unif}}) = 1$, no matter the cardinality of $A$.

Recall that $\mu^{\kappa}_j$ is the probability measure associated with $\pi_j(\kappa | \bm{X}_j)$. As $\mu^{\kappa}_j$ gets closer and closer to the discrete uniform distribution--expressing complete ambivalence as to the location of a changepoint---$H(\mu^{\kappa}_j)$ approaches 1. Therefore, the practitioner may set some threshold $\eta$ for the normalized entropy and remove $\hat{\kappa}_j$ as a changepoint candidate if $H(\mu^{\kappa}_j) \geq \eta.$ 
We take as our final changepoint estimates those with normalized entropy less than $\eta,$ meaning that the posterior distributions are reasonably concentrated. Selection of $\eta$ can be done by appealing to the normalized entropy of a reference distribution---such as the binomial---but we leave this calibration as future work.

Estimating many changepoints and then eliminating extraneous ones based on large (or small) values of some criterion is known as ``bottom-up segmentation'' and has been shown to outperform binary segmentation on many different datasets \citep{keogh2001, bwd}. A generic description of bottom-up segmentation can be found in \citet[Algorithm 5]{selective_review}. Our approach is a variation on that version of bottom-up segmentation. 

To utilize all the data between estimated changepoints in our final analysis (e.g. to provide lower variance estimates of $\bmb$) we can instead, or additionally, perform bottom-up segmentation for $\kappa_j$ based on the posterior distributions $\pi_j(\kappa \mid \bm{X}_j)$ generated during the warm-up period described in Section~\ref{ss:warm_up}. It can be beneficial to employ a second warm-up period as well, with a lower entropy threshold---see Section S6.2 in the Supplementary Material. We implement this entire procedure (two warm-up periods and then a final fit with resulting segments) as the default in the code that accompanies this article, which is available at \url{https://github.com/manilasoldier/bclr}.

\section{Conclusion}

In this article, we have presented a Bayesian changepoint method that utilizes logistic regression to detect changes in an interpretable and parsimonious fashion while imposing few assumptions on the data-generating process. Our method also learns the nature of the changepoint and provides uncertainty quantification on both the location of change and the coordinates in which the change occurs. We have also provided a canonical topological feature embedding for detecting changes in image series that outperforms standard Euclidean features and have demonstrated our method's competitive performance on a variety of tasks. 

The Bayesian changepoint method introduced in this article could be extended in a number of different directions: 

\vspace{5pt}
\noindent \textbf{Alternative classifiers.} We could consider alternatives to the combination of a logistic regression model with a multivariate normal prior. For example, we expect that choosing a prior distribution that induces sparsity in the regression coefficients \citep{horseshoe, spike_slab, duan2023} would lead to improved inferences and greater interpretability for high-dimensional time series. To capture complex changes with less feature engineering (at the expense of some interpretability), we could replace the logistic regression model with a more flexible classifier based on Bayesian additive regression trees \citep{bart} or kernel methods \citep{zhu2005, shawe2004}. 

\vspace{5pt}
\noindent \textbf{More complex data types.} As we have emphasized through the article, the proposed Bayesian changepoint method treats the observed time series as a sequence of covariate vectors. As a result, there is no need to specify a model for the data, and the method can be applied to data of mixed type. For these same reasons, we expect the method could be extended in a conceptually straightforward way to handle more complex data types such as time-indexed network, functional, or shape data.

\vspace{5pt}
\noindent \textbf{Multiple changepoints.}  
Another, possibly more natural, approach to extending our method to the multiple changepoint setting would be to allow the latent variables $Y_1, \ldots, Y_n$ to take values in the set $\{1, 2, \ldots, L+1\}$ where $L$ is the number of changepoints. In that case, the logistic regression model could be replaced with a multinomial logistic regression model. The quasi-likelihood for this setup is described in Section S6.1 of the Supplementary Material. This extension is appealing from both a conceptual and a practical perspective, but it raises distinct computational and methodological challenges that warrant a separate treatment. Another interesting direction would be to conduct posterior inference (or some approximation thereof) over all possible changepoint configurations by placing a prior on the space of compositions of the set $\{1, \dots, n\},$ as in \citet{martinez_mena} or \citet{jin2022bayesian}.  



\vspace{5pt}

\noindent We intend to pursue these investigations and extensions in future work.



\vspace{5pt}

\bibliographystyle{plainnat}
\bibliography{ChangepointTDA_V3}

\end{document}


\title{\textbf{Supplementary Material for \\ ``Bayesian changepoint detection via logistic regression \\ and the topological analysis of image series''}}
\author{Andrew M. Thomas, Michael Jauch, and David S. Matteson}
\date{}

\maketitle

\section{Additional theory}

In this section we expand on some of the ideas of Section 2 of the main document. 

\subsection{The intercept term and centering} \label{ss:int_norm}

Here we elaborate on some of the data preprocessing steps we choose to employ. As mentioned in the main document, in our experiments, we found that without omitting the intercept and centering our data, our changepoint estimates would tend concentrate at $\kappa=1$ or $\kappa=n-1$. We first examine why we omitted the intercept. Consider the full conditional distribution $\pi(\kappa \mid \bmb, \x)$, which is proportional to 
$$
\exp\Big\{\sum_{i=\kappa+1}^n \bm{x}_i^{\top}\bmb\Big \}\pi(\kappa);
$$
If we include an intercept term $\beta_0$ in our model, then our posterior becomes
\begin{align*}
\exp\Big\{\sum_{i=\kappa+1}^n (\beta_0 + \bm{x}_i^{\top}\bmb)\Big\}\pi(\kappa) = \exp\Big\{\sum_{i=\kappa+1}^n \bm{x}_i^{\top}\bmb\Big \}e^{(n-\kappa)\beta_0}\pi(\kappa).
\end{align*}
Thus, including an intercept term is equivalent to imposing a prior of $e^{(n-\kappa)\beta_0}\pi(\kappa)$ on there being a changepoint at $\kappa$. If we suppose that $\pi(\kappa) \propto 1$ so that we have a discrete uniform prior on $\kappa$. Then any nonzero $\beta_0$ will bias  posterior probability that $\kappa = n-1$ (changepoint at the end) or that $\kappa = 1$ (an immediate changepoint) towards 1. As we do not wish to bias our posterior of $\kappa$ in this way, we omit an intercept term.

Now that we have justified omitting the intercept, let us consider our data $\x$. Assume for simplicity that we are in the uninformative setup where we place a discrete uniform prior on $\kappa$. To discover the changepoint (conditional on $\bmb$) we want the true changepoint to satisfy $\hat{\kappa} = \kappa^*$, where 
$$
\hat{\kappa} = \max_{\kappa} \pi(\kappa \mid \bmb, \x) = \max_{\kappa} \exp\Big\{\sum_{i={\kappa+1}}^n \bm{x}_i^{\top}\bmb \Big\}.
$$
This will occur if 
\begin{equation}\label{e:crit_max}
\frac{\exp\Big\{\sum_{i={\hat{\kappa}+1}}^n \bm{x}_i^{\top}\bmb \Big\}}{\exp\Big\{\sum_{i=j+1}^n \bm{x}_i^{\top}\bmb \Big\}} \geq 1
\end{equation}
for all $j \neq \hat{\kappa}$. The equation \eqref{e:crit_max} is equivalent (for all $j \neq \hat{\kappa}$) to
\begin{align}
&\sum_{i={\hat{\kappa}+1}}^n \bm{x}_i^{\top}\bmb \geq \sum_{i={j+1}}^n \bm{x}_i^{\top}\bmb, \notag \\
\Leftrightarrow &
\begin{cases}
\frac{1}{|\hat{\kappa}-j|}\sum_{i=j+1}^{\hat{\kappa}} \bm{x}_i^{\top}\bmb \leq 0 &\mbox { if } j < \hat{\kappa}  \\
\frac{1}{|\hat{\kappa}-j|}\sum_{i=\hat{\kappa}+1}^{j} \bm{x}_i^{\top}\bmb \geq 0&\mbox { if } j > \hat{\kappa} .
\label{e:how}
\end{cases}
\end{align}
If by, via a random choice of $\bmb$, it happens that $\bm{x}_i^{\top}\bmb > 0$ for all $i = 1, \dots, n$ then we have $\hat{\kappa}=1$. Similarly if $\bm{x}_i^{\top}\bmb < 0$ for all $i$, it must be the case that $\hat{\kappa}=n-1$. Because we do not want to bias our method with an intercept, we must choose another type of solution. 

A reasonable desideratum for any changepoint method is that it should perform well on a piecewise constant signal. Suppose that $\bm{x}_i^{\top}\bmb = a$ for $i \leq \kappa^*$ and $\bm{x}_i^{\top}\bmb = b$ for $i > \kappa^*$. Then we have
$$
\bar{\bm{x}}^{\top}\bmb = (\kappa^* a + (n-\kappa^*)b)/n > a.
$$
If we define $\bm{z}_i :=\bm{x}_i^{\top}\bmb$ and then center $\bm{z}_i$, the condition \eqref{e:how} becomes 
\begin{equation*}
\begin{cases}
\frac{1}{|\hat{\kappa}-j|}\sum_{i=j+1}^{\hat{\kappa}} \bm{z}_i \leq \bar{\bm{z}} &\mbox { if } j < \hat{\kappa}  \\
\frac{1}{|\hat{\kappa}-j|}\sum_{i=\hat{\kappa}+1}^{j} \bm{z}_i \geq \bar{\bm{z}} &\mbox { if } j > \hat{\kappa}, \label{e:centered}
\end{cases}
\end{equation*}
for all $j \neq \hat{\kappa}$, which is satisfied uniquely for $\hat{\kappa} = \kappa^*$. However, for a general signal we also have
\begin{align*}
\bm{z}_i - \bar{\bm{z}} &= \bm{x}_i^{\top}\bmb - \frac{1}{n} \sum_{j=1}^n \bm{x}_j^{\top}\bmb \\
&= \bm{x}_i^{\top}\bmb - \bm{\bar{x}}^{\top}\bmb \\
&= (\bm{x}_i^{\top} - \bm{\bar{x}})^{\top}\bmb,
\end{align*}
so it suffices to center the $\bm{x}_i$ to achieve the same centering of the $\bm{z}_i$. Other means of centering could in principle be used but the sample mean has the desirable property that $\bar{\bm{z}} = \bar{\bm{x}}^{\top}\bmb$. As further support in favor of these preprocessing steps, centered observations also appear in the one-sided likelihood-ratio test statistic of a change in mean, if $\bm{x}_i^{\top}\bmb$ are assumed independent Gaussian with constant variance (cf. p. 10, \cite{chen2012}, for example). Finally, a centered constant sequence (i.e. a constant sequence at zero) will yield a uniform marginal posterior $\pi(\kappa \mid \x) = 1/(n-1)$ for $\kappa = 1, \dots, n-1$. 

\subsection{Representation of the marginal posterior}

The P{\'o}lya-Gamma approach furnishes a representation of $\pi(\kappa \mid \x)$, which is potentially interesting from the standpoint of how BCLR specifies the marginal posterior for $\kappa$. Assume a discrete uniform prior for $\kappa$. We can see that
\begin{align*}
\pi(\kappa \mid \x) &\propto \int \pi(\kappa , \bmb \mid \x) \dif{\bmb} \\
&= \int \mathcal{Q}(\bmb, \kappa \mid \x)\pi(\bmb)\dif{\bmb}.
\end{align*}
For reference (see \citealp{polson2013}), a P{\'o}lya-Gamma $\text{PG}(1,0)$ random variable has density
\[
p(\omega) = \sum_{\ell=0}^{\infty} (-1)^{\ell}\frac{2\ell+1}{\sqrt{2\pi\omega^3}}e^{-(2\ell+1)^2/(8\omega)}, \quad \omega \geq 0,
\]
and is equal to an infinite convolution of Gamma distributions. We may now state the result.

\begin{proposition}\label{p:pikx}
Let $\bmb$ have a mean zero multivariate normal prior with positive semi-definite covariance matrix $\Sigma$. Then 
\[
\pi(\kappa \mid \x) \propto \int_{[0,\infty)^n} \det(\bm{V}_\omega)^{1/2} \exp\Big(  \vec{\bm{x}}_{\kappa}^{\top}\bm{V}_{\omega}\vec{\bm{x}}_{\kappa}/2\Big) \prod_{i=1}^n p(\omega_i) \dif{\omega_i},
\]
where we recall that $\bm{V}_\omega = (\BX^{\top}\Omega\BX + \Sigma^{-1})^{-1}$, $\Omega = \mathrm{diag}(\bm{\omega})$ and define 
\[
\vec{\bm{x}}_{\kappa} := \sum_{i=1}^n (-1)^{\mathbf{1}\{i \leq \kappa\}}\bm{x}_i/2.
\]
\end{proposition}

\begin{proof}
First, let us define
\[
(\bm{z}_{\kappa})_i := \frac{(-1)^{\mathbf{1}\{i \leq \kappa\}}}{2\sqrt{\omega_i}},
\]
set $\BX_\omega$ to be the matrix with row $i$ equal to $\sqrt{\omega_i} \bm{x}_i^{\top}$. Also define $a_i^{\kappa} := \text{sign}(i-\kappa)/2$ for $\kappa=1, \dots, n$ (where $\text{sign}(0)=-1$). By Theorem 1 in \cite{polson2013}, we have that 
\begin{align*}
\pi(\kappa \mid \x) &\propto \int_{\R^d} \int_{0}^{\infty} \cdots \int_0^{\infty} \prod_{i=1}^n \bigg[e^{a_i^{\kappa}\bm{x}_i^{\top}\bmb}e^{-\omega_i (\bm{x}_i^{\top}\bmb)^2/2} p(\omega_i) \bigg] \pi(\bmb) \dif{\bm{\omega}} \dif{\bmb} \\
&= \int_{\R^d} \int_{0}^{\infty} \cdots \int_0^{\infty}  \exp\bigg\{\sum_{i=1}^n a_i^{\kappa}\bm{x}_i^{\top}\bmb -\omega_i (\bm{x}_i^{\top}\bmb)^2/2\bigg\} p(\omega_1) \cdots p(\omega_n) \pi(\bmb) \dif{\bm{\omega}} \dif{\bmb}.
\end{align*}
Completing the square yields
\begin{align}
a_i^{\kappa}\bm{x}_i^{\top}\bmb -\omega_i (\bm{x}_i^{\top}\bmb)^2/2 = -\frac{1}{2}\big[(\sqrt{\omega_i}\bm{x}_i)^{\top}\bmb - a_i^{\kappa}/\sqrt{\omega_i}\big]^2 + \frac{1}{8\omega_i}, \label{e:8om}
\end{align}
as $(a_i^{\kappa})^2 = 1/4$. Therefore, it remains to study the behavior of
\begin{align*}
&\exp\Bigg\{-\frac{1}{2}\bigg( \sum_{i=1}^n \big[(\sqrt{\omega_i}\bm{x}_i)^{\top}\bmb - a_i^{\kappa}/\sqrt{\omega_i}\big]^2 + \bmb^{\top}\Sigma^{-1}\bmb \bigg) \Bigg\} \\
&\qquad= \exp\Bigg\{-\frac{1}{2} \Big[ (\bm{z}_{\kappa}-\BX_{\omega}\bmb)^{\top}(\bm{z}_{\kappa}-\BX_{\omega}\bmb) + \bmb^{\top}\Sigma^{-1}\bmb \Big] \Bigg\},
\end{align*}
upon integrating out $\bmb$. However, by standard linear model results (cf. Theorem 3.7 in \citealp{seber_lee}), we have that 
\begin{align*}
&(\bm{z}_{\kappa}-\BX_{\omega}\bmb)^{\top}(\bm{z}_{\kappa}-\BX_{\omega}\bmb) + \bmb^{\top}\Sigma^{-1}\bmb \\
&\qquad= (\bmb^{\top}-\bm{m}_{\omega})\bm{V}_{\omega}^{-1}(\bmb-\bm{m}_{\omega}) + \bm{z}_{\kappa}^{\top}(I_n + \BX_{\omega}\Sigma\BX_{\omega}^{\top})^{-1}\bm{z}_{\kappa},
\end{align*}
where $\bm{m}_{\omega}$ is as defined in the main document. Integrating out $\bmb$---which produces $\det(\bm{V}_{\omega})^{1/2}$---leaves only the term $-(1/2)\bm{z}_{\kappa}^{\top}(I_n + \BX_{\omega}\Sigma\BX_{\omega}^{\top})^{-1}\bm{z}_{\kappa}$ to deal with. Here we use the Sherman-Morrison-Woodbury identity, which asserts that if $A, C$ and $C^{-1} + DA^{-1}B$ are nonsingular then 
$$
(A+BCD)^{-1} = A^{-1} - A^{-1}B(C^{-1} + DA^{-1}B)^{-1}DA^{-1}.
$$
Setting $A = I_n$, $B = \x_{\omega}$, $C=\Sigma$, and $D = \x_{\omega}^{\top}$, the matrices satisfy the requirements and we get that 
$$
(I_n + \BX_{\omega}\Sigma\BX_{\omega}^{\top})^{-1} = I_n - \x_{\omega}\bm{V}_{\omega}\x_{\omega}^{\top}.
$$
As $-(1/2)\bm{z}_{\kappa}^{\top}\bm{z}_{\kappa} = -\sum_{i=1}^n \frac{1}{8\omega_i}$,
this cancels out the term seen in \eqref{e:8om}. Thus we are left with only the term 
$$
(1/2)\bm{z}_{\kappa}^{\top}\x_{\omega}\bm{V}_{\omega}\x_{\omega}^{\top}\bm{z}_{\kappa},
$$
which yields our final result for the exponent term based off the definition of $\vec{\bm{x}}_{\kappa}$.
\end{proof}

\section{Additional details on simulation}

For the simulations, we examined 5 different methods inclusive of BCLR. We have chosen the default settings for all algorithms with two minor exceptions: we set the minimum number of observations between changepoints to 2 for ECP and KCP, and we ran the KCP method with the radial basis function kernel. For CF, we used the \emph{changeforest} function in the same package with the classifier chosen to be random forests and the multiple changepoint method chosen to be binary segmentation. The bandwidth for the RBF kernel for KCP was chosen via the ``median heuristic'' \citep{median_heur}. Note that we use the dynamic programming formulation for changepoint detection in the \emph{ruptures} package for KCP, and assume that there is one changepoint in each of the time series we consider. As mentioned in the main document, we have developed a Python package BCLR to accompany this paper. The BCLR Python package is available at \url{https://github.com/manilasoldier/bclr}. The information and code for the simulations can be found in the ``experiments'' folder in the same location. 

We also considered the Bayesian hierarchical method BHM as developed in \cite{jin2022bayesian}. For the BHM method, we tailored the code available with the article (\url{https://www.tandfonline.com/doi/suppl/10.1080/00401706.2021.1927848}) to our context, in particular setting the window parameter to $m_I = 2$. For the first simulated image series $V^{(1)}$, the BHM method returned a changepoint at $\kappa=21$ (it does slightly better when using unstandardized TDA features, recovering a changepoint at $\kappa=24$). However, we did not include BHM in our main document analyses as it did not produce satisfactory results. 

In the main article we considered quantile intervals for our changepoint estimates. We can also define a $(1-\alpha)\times 100\%$ credible set with the highest probability mass to be
\[
C_{\alpha, \ell} := \argmin_{S_\alpha} | S_\alpha |
\]
where $S_{\alpha}$ are subsets of $1, \dots, n-1$ satisfying
$$
\sum_{\kappa \in S_{\alpha}} p_\ell(\kappa) \geq 1-\alpha, \text{ and } p_\ell(\kappa) \geq p_{\ell}(\kappa') \text{ for } \kappa' \not \in S_\alpha.
$$
Note that in case $C_{\alpha, \ell}$ is not unique, we take $C_{\alpha, \ell}$ with the smallest sum. The highest posterior mass intervals just described can be seen for Experiment 1 from the main document in Table~\ref{t:cov_cred2} and for Experiment 2 in Table~\ref{t:cov_cred3}. On a computational note, we used both \texttt{R} and \texttt{Python} for our analyses. We used the \texttt{BayesLogit} package \citep{polson2013} to sample from the P{\'o}lya-Gamma distribution in \texttt{R} and the \texttt{polyagamma}\footnote{https://pypi.org/project/polyagamma/.} package in \texttt{Python}.

\begin{table}[t]
\centering
\bgroup
\def\arraystretch{1.25}
\begin{tabular}{l|rrrrr}
& \multicolumn{5}{c}{$\P(\kappa^* \in C_{\alpha})$} \\
$(1-\alpha)\times100\%$ & 50\% & 80\% & 90\% & 95\% & 99\% \\
\hline \hline
TDA & 0.749 & 0.887 & 0.939 & 0.962 & 0.991 \\
PCA & 0.168 & 0.32 & 0.432 & 0.511 & 0.659 \\
\end{tabular}
\egroup
\caption{Estimated credible interval coverage probabilities for $\kappa^*=25$ for our BCLR method using the credible set calculated from values of highest posterior mass.}
\label{t:cov_cred2}
\end{table}


\begin{table}[t]
\centering
\bgroup
\def\arraystretch{1.25}
\begin{tabular}{l|rrrrr}
$(1-\alpha)\times100\%$ & 50\% & 80\% & 90\% & 95\% & 99\% \\
\hline \hline
$\P(\kappa^* \in I_{\alpha})$ & 0.429 & 0.55 & 0.632 & 0.683 & 0.756 \\
$\P(\kappa^* \in C_{\alpha})$ & 0.4 & 0.535 & 0.609 & 0.667 & 0.746 \\
\end{tabular}
\egroup
\caption{Estimated credible interval coverage probabilities for $\kappa^*=40$ for our BCLR method (with simple prior, i.e. \texttt{bclr-1}) using the TDA features for both methods of constructing a credible set/interval seen in this article.}
\label{t:cov_cred3}
\end{table}


\subsection{Other priors in Experiment 2} \label{ss:other}

We ran Experiment 2 with a variety of different priors for $(\kappa, \bmb)$, which we recount here. The results of these experiments (run on the same images as the rest of Experiment 2) are detailed in Table~\ref{t:exp2_priors}. For $\bmb$ we considered priors of $N(0, 3I_d)$ and the ``data-driven prior'' of the main text. For $\kappa$ we considered three separate priors. First was the discrete uniform prior on $\{1, \dots, 49\}$. Second was the ``binomial prior'' discussed in the main text. The last prior used was an interpolation of the uniform and binomial distribution (``uni-binom'') specified by
$$
\pi(\kappa) \propto \Bigg[\dbinom{48}{\kappa-1}(0.8)^{\kappa-1}(0.2)^{48-\kappa+1}\Bigg]^{\nu}, \quad \kappa = 1, \dots, 49
$$
with $\nu = 0.02$, so that the prior mode of this uniform/binomial distribution has unique mode at $\kappa=40$. The method \texttt{bclr-1} used a $N(0, 3I_d)$ prior for $\bmb$ and a uniform prior for $\kappa$. The variant \texttt{bclr-2} had the same prior for $\bmb$ and the uni-binom prior for $\kappa$; \texttt{bclr-3} used the data-driven prior for $\bmb$ and the uni-binom prior for $\kappa$. Finally \texttt{bclr-4} used the $N(0, 3I_d)$ prior for $\bmb$ and the binomial prior for $\kappa$. From Table~\ref{t:exp2_priors} and Table 3 in the main document, it seems like an informative choice of $\bmb$ is more important than a rather informative $\kappa$, though this should warrant further investigation.

\begin{table}[t]
\centering
\bgroup
\def\arraystretch{1.25}
\begin{tabular}{l|rrrr}
Method & \texttt{bclr-1} & \texttt{bclr-2} & \texttt{bclr-3} & \texttt{bclr-4}  \\
\hline \hline
\% Exact & 0.334 & 0.345 & 0.516 & 0.432 \\
& (0.015) & (0.015) & (0.016) & (0.016) \\
\hline 
RMSE & 4.498 & 4.436 & 2.490 & 2.311 \\
& (4.649) & (4.483) & (2.299) & (2.000) \\
\hline \hline
\end{tabular}
\egroup
\caption{\% Exact and RMSE for 1000 simulated videos for 4 different priors for BCLR in the case of $\kappa^*=40$ and $X_{ijk} = Z_{ijk} + 1$. The nomenclature for \texttt{bclr-i} for \texttt{i}=$1,2,3,4$ can be seen in Section~\ref{ss:other}. Standard errors for \% Exact and RMSE indicated in parentheses.}
\label{t:exp2_priors}
\end{table}

\subsection{Quality of convergence of Gibbs sampler}

\begin{figure}[t]
\centering
\includegraphics[width=0.85\textwidth]{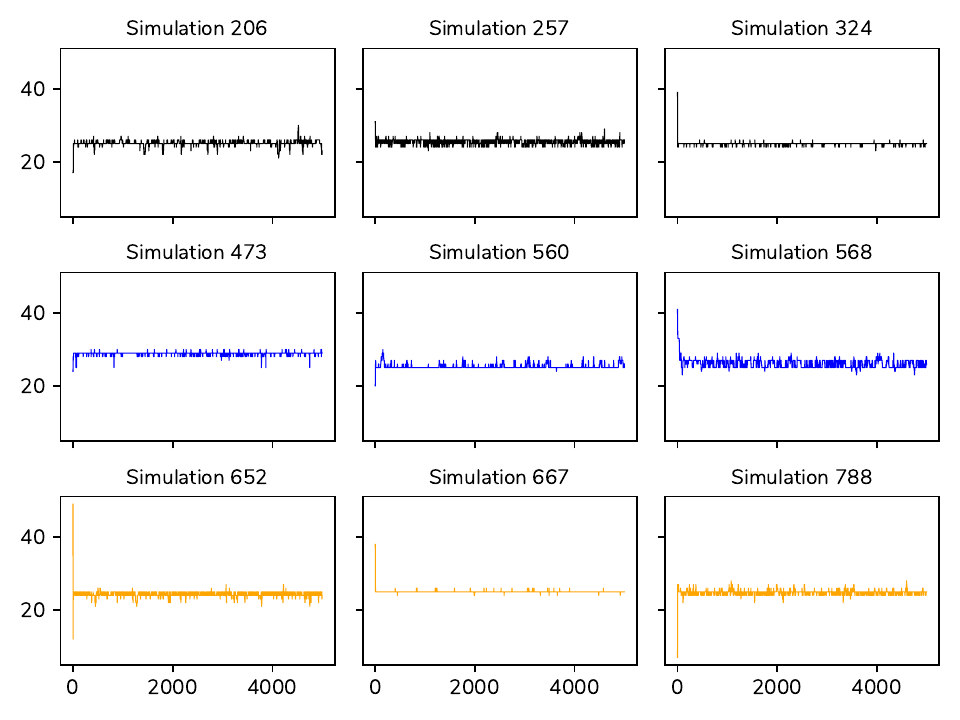}
\caption{Traceplots of 9 randomly chosen simulations when $\kappa^*=25$.}
\label{f:trace25}
\end{figure}

\begin{figure}[h]
\centering
\includegraphics[width=0.85\textwidth]{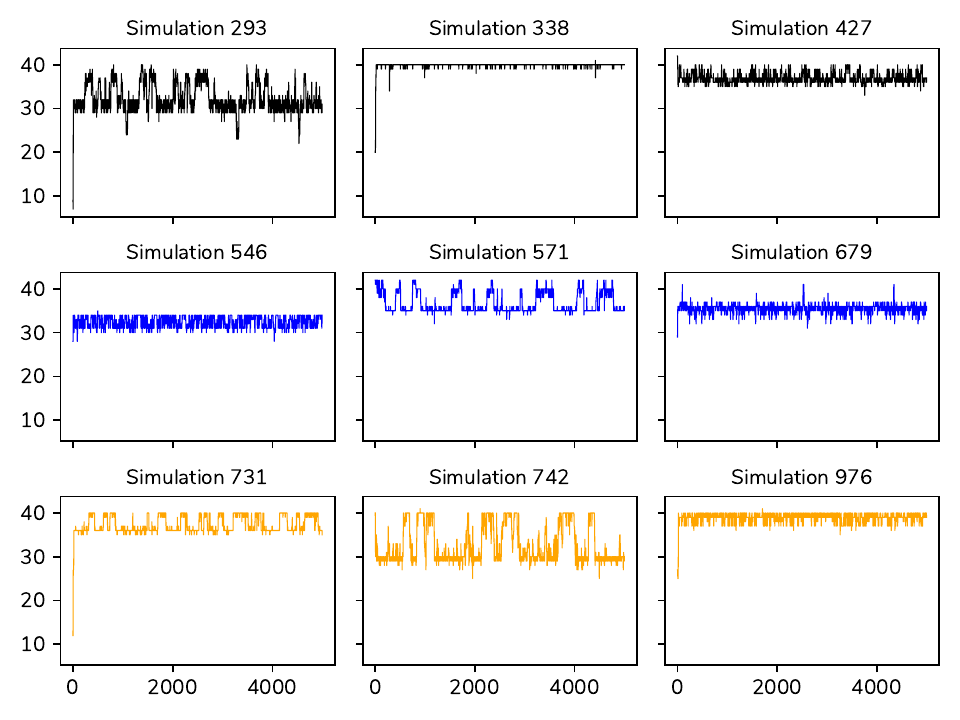}
\caption{Traceplots of 9 randomly chosen simulations when $\kappa^*=40$.}
\label{f:trace40}
\end{figure}

To assess convergence of the Gibbs sampler in the simulations, we provide traceplots of $\kappa$ for a few simulations in Figure~\ref{f:trace25} (when $\kappa^*=25$) and Figure~\ref{f:trace40} (when $\kappa^*=40$, using the ``default'' \texttt{bclr-1} method mentioned in Table~\ref{t:exp2_priors}). Convergence is rapid in the case of $\kappa^*=25$ (and stronger signal). The convergence in the case of $\kappa^*=40$ (and weaker signal) is more of a mixed bag.  It would seem to indicate that our algorithm in indecisive as indicated by the Gibbs sampler. However, in each of the cases we see in Figure~\ref{f:trace40}, there are plateaus around $\kappa^* = 40$, even if this is not the posterior mode. The frequent switching between the ``correct'' and a much lower $\kappa$ seems to indicate these chains evince reasonable mixing properties, even though the posterior mode yields an incorrect changepoint estimate. 

\subsection{Timing}

For a single image series of 50 frames using the 36-dimensional topological features (with changepoint at $\kappa^*=25$), the mean computation time for 5000 Monte Carlo iterations was 4.93 (standard deviation = 0.85) seconds over 100 separate simulations with garbage collection, using the Python $\texttt{timeit}$ module. The minimum computation time was 4.03 seconds. This timing experiment was implemented on a MacBook Air with an Apple M3 8-Core CPU with 16GB of RAM, though Monte Carlo iterations were all computed serially. 
As a final note, it is likely that using a binary probit model may be computationally quicker here, as we would not have to sample from the P{\'o}lya-Gamma distribution. Based on the previous subsection, a lower number of Monte Carlo iterations may suffice to get reasonable results as well. 

\subsection{Posterior mean as the changepoint estimate}

Throughout the article, we used the posterior mode as a changepoint estimate. However, we could have also used the posterior mean. In Section 4.1 we mentioned that we also calculated the RMSE of the 1000 simulations in Experiments 1 and 2 using the posterior mean within each simulation. We will call this $RMSE_1$. We will denote the RMSE reported in the main document as $RMSE_0$. Instead of calculating the RMSE within each posterior sample and averaging, we may take the posterior mode within each of the 1000 samples and then calculate the RMSE (this is more akin to what was done with the other methods). We call denote this as $RMSE_2$.

For Experiment 1, the RMSE (from $\kappa^*=25)$ for the posterior mean ($RMSE_1$) was 0.926, which was fairly close to the reported value of $RMSE_0 = 0.948$. The RMSE for the posterior mode ($RMSE_2$) was 1.072. In Experiment 2, for \texttt{bclr-*}, $RMSE_1=1.403$ (from $\kappa^*=40$, for the posterior mean), and $RMSE_2=1.489$. However, for \texttt{bclr-1} $RMSE_1=6.140$ and $RMSE_2=6.465$---much higher than $RMSE_0$ seen in Table~\ref{t:exp2_priors}. Jensen's inequality shows that for $\kappa_{\ell j} \sim \pi(\kappa \mid \x_\ell)$ (as in the posterior sample for a single simulated video) with $\ell = 1, \dots, n$ and $j=1, \dots, m$ we have that 
$$
RMSE_0 = \frac{1}{n} \sum_{\ell=1}^n \sqrt{\frac{1}{m} \sum_{j=1}^m (\kappa_{\ell j} - \kappa^*)^2 } \leq \sqrt{\frac{1}{n} \sum_{\ell=1}^n (\bar{\kappa}_\ell - \kappa^*)^2  + \frac{1}{nm} \sum_{\ell=1}^n \sum_{j=1}^m (\kappa_{\ell j} - \bar{\kappa}_\ell)^2 },
$$
where $\bar{\kappa}_\ell = (1/m)\sum_{j=1}^m X_{\ell j}$. Note that in Experiments 1 and 2 we have $m=2500$ and $n=1000$. This indicates that if our posterior distributions concentrate around the mean for each simulation we will have $RMSE_0 \leq RMSE_1$ where
$$
RMSE_1 = \sqrt{\frac{1}{n} \sum_{\ell=1}^n (\bar{\kappa}_\ell - \kappa^*)^2}.
$$
This helps explain the disparity in values of $RMSE_0$ and $RMSE_1$, $RMSE_2$ for the case of \texttt{bclr-1} (see also the concentration phenomenon in Figure~\ref{f:trace40})\footnote{For completeness,
$RMSE_2 = \sqrt{\frac{1}{n} \sum_{\ell=1}^n (\hat{\kappa}_\ell - \kappa^*)^2}$, where $\hat{\kappa}_\ell = \inf \{\argmax_{\kappa} p_\ell(\kappa)\}$, using the notation from the main document.}. 

In many cases value of $RMSE_0$ is smaller than the RMSE for the other methods. This means that the average standard deviation of an individual posterior sample $\kappa_{\ell j}$ is less than the standard deviation of the changepoint estimates of the other methods. 

\section{Further insight into interval estimation}

For simplicity we ignore the feature embedding $\psi$ in the following. In our simulations, we generate data $\bm{X}$ according to a change occurring at frame $\kappa^*$, which we state to have likelihood $p(\bm{X} \mid \kappa^*)$. We then calculate the posterior distribution of our changepoint based off of $\pi(\kappa \mid \bm{X})$. The probability of the credible interval containing $\kappa^*$ that we approximate in the paper is then 
\[
\int_{(\R^d)^n} \ind{ q_{\alpha/2}(\bm{X}) \leq \kappa^* \leq q_{1-\alpha/2}(\bm{X})} p(\bm{X} \mid \kappa^*) \dif{\bm{X}}.
\]
where 
\[
q_{\alpha}(\bm{X}) := \inf\{\kappa: \sum_{i=1}^{\kappa} \pi(\kappa \mid \bm{X}) \geq \alpha\}.
\]
It is straightforward to see that if the data (especially under transformation, or if the noise corrupts the signal) does not contain good information about the change point, then there is little reason to believe that posterior will capture the change. There is also the secondary source of error due to the approximation of $\pi(\kappa \mid \bm{X}).$


\section{Persistence images}

In this section we demonstrate the use of persistence images as one of our embeddings $f$ to 1) extend the results of \cite{obayashi2018}, and 2) try to capture regions of interest within images (that correspond to change). We do this for the first simulated video $V^{(1)}$ where $\kappa^* = 25$. In contrast to $f_{\text{stat}}$ above $f_{\text{PI}}$ is tailored to the aforementioned objectives, not to the objective of finding the most appropriate univariate representation of topological change. Before we define $f_{\text{PI}}$ we must briefly define the persistence image.

\begin{definition}[Definition 1 in \citealp{adams2017}] \label{d:pi}
Fix a persistence diagram $\mathcal{D}^k$ (here $k = 0,1$). A \emph{persistence image} is a discrete sampling of the function defined on $\R \times [0, \infty)$ by 
\[
\rho(x,y) = \sum_{p \in M} w(b_p, l_p) \exp\Big(-\frac{(b_p - x)^2 + (l_p - y)^2}{2\sigma^2} \Big),
\]
called the \emph{persistence surface}, where $w: \R \times [0, \infty) \to [0, \infty)$ is a weight function that is continuous, piecewise differentiable and satisfies $w(\cdot, 0) = 0$.
\end{definition}

The index set $M$ in the case of our setting of cubical homology is the set of local minima of the image. In Definition~\ref{d:pi} we see that the persistence image can be calculated for either $\mathcal{D}^0$ or $\mathcal{D}^1$. In particular, we could calculate it for both and append the vectors together to create $f_{\text{PI}}$. For $V^{(1)}$, which we analyze here, the $0^{th}$ homology is most relevant so we will simply take the persistence image of $\mathcal{D}^0$ and define 
\[
\big(f_{\text{PI}}\big)_{ij} = \rho(x_i, y_j), \quad 1\leq i,j \leq 6,
\]
where $x_i$ and $y_j$ are sampled at 6 equally spaced locations in the intervals $[\min_{p} b_p, \max_{p} b_p]$ and $[\min_p l_p, \max_p l_p]$ respectively (with the endpoints included). That is, $(x_i, y_j) \in B$ where 
\begin{equation}\label{e:B}
B := \{(\min_{p} b_p + kr,  \min_p l_p + ms): m, k = 0,1,2,3,4,5 \}
\end{equation}
where $r = (\max_{p} b_p-\min_{p} b_p)/5$ and $s = (\max_p l_p-\min_p l_p)/5$. The summary $f_{\text{PI}}$ is flattened to a vector in $\R^{36}$ by concatenating each column. Of course, we could make this persistence image to have arbitrary size, but choose $6 \times 6$ for symmetry and because there is relatively little data (only 50 frames in our image series). 

\begin{figure}[t]
\includegraphics[width=\textwidth]{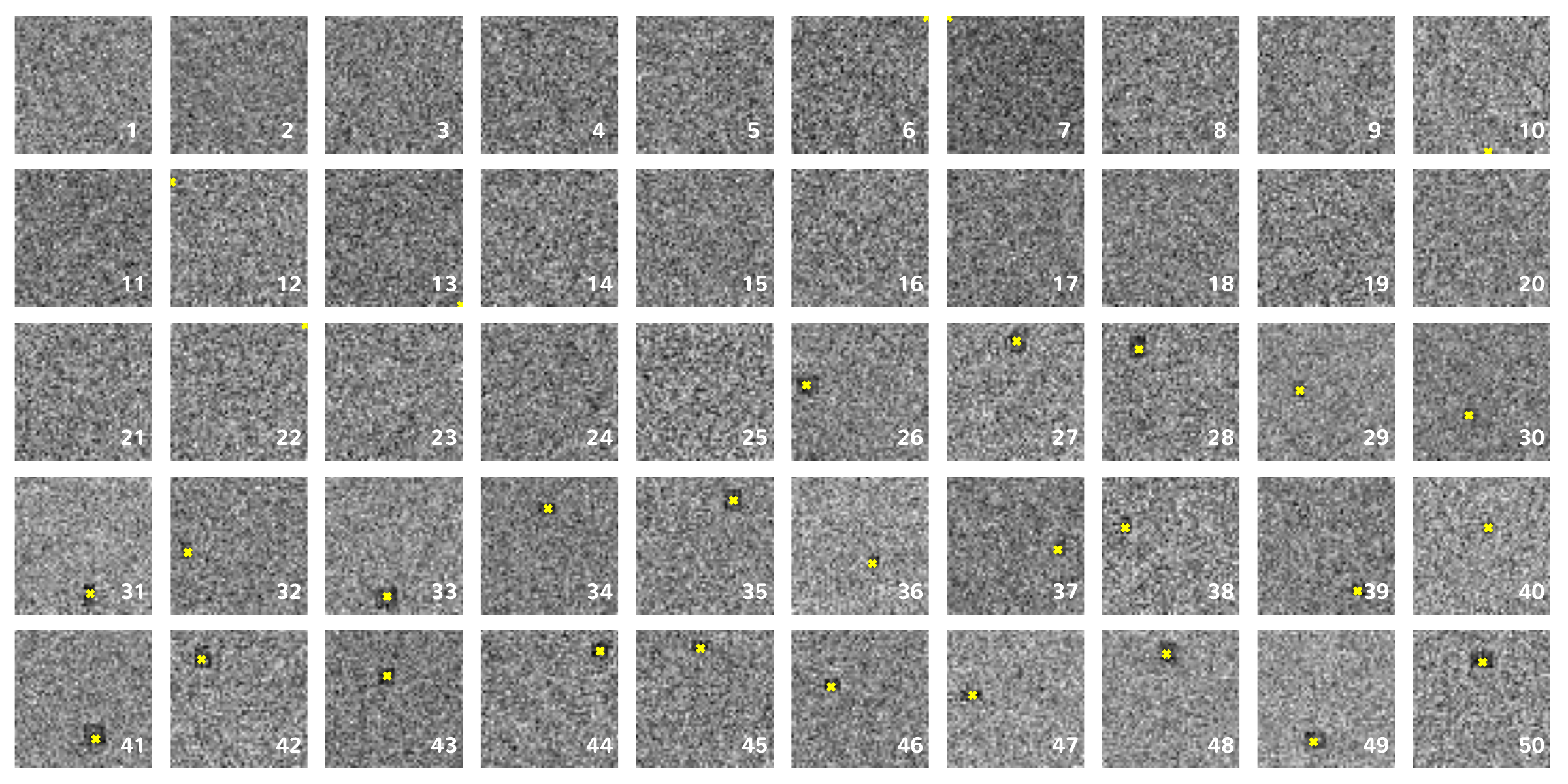}
\caption{All 50 frames of the demonstration video with $\kappa^*=25$, with pixels corresponding to regions within persistence images where $\beta$ coefficients were large (i.e. positively associated with a change having occurred). }
\label{f:all50pi}
\end{figure}
Some spurious features are seen in various frames (frames 6, 7, 10, 12, 13, and 22) of Figure~\ref{f:all50pi} on the edges of the images. This could easily be fixed by restricting our analysis to a subimage away from the image boundaries. 

We use the inverse tangent weight function described in \cite{obayashi2018} with $C=0.5$ and $p=1$. After $\bmb$ coefficients are esimated for the vector $f_{\text{PI}}$, we need some way to map said coefficients back to the image. As mentioned in Section S5.3, each point of $\mathcal{D}^0$ maps injectively to a pixel in the image $X_i$. Let $\pi_{ij}(\bmb \mid \bm{X})$, $1 \leq i,j \leq 6$ be the posterior distribution for the coefficient associated to $(f_{\text{PI}})_{ij}$ from running BCLR. As our data is standardized, there is a meaningful zero and hence we may select influential coefficients by choosing $(i,j)$ such that the $\alpha$ (alternatively $1-\alpha$) quantile of $\pi_{ij}(\bmb \mid \bm{X})$ is greater than (less than) 0 for small $\alpha > 0$. From here we may uniquely associate such $(i,j)$ with a coordinate in $B$ from equation \eqref{e:B}---Specifically the point in $B$ with $k = (i-1)$ and $m = (j-1)$.

For each point $(b, l)$ in the transformed persistence diagram, we check to see if this is next to a ``significant'' coordinate $(b_0, l_0) \in B$, by checking if $(b,l)$ lies within the Voronoi cell of $(b_0, l_0)$, i.e. 
\[
\max\{ 2|b-b_0|/r, 2|l-l_0|/s \} \leq 1.
\]
If so, we then display its location---see Figure~\ref{f:all50pi}. Though we only display the local minima that are positively associated with a change, we could also display those which are negatively associated with a change, in the same manner. For this image we calibrated $\alpha=0.45$, which yielded the best results. In practice, the quantile $\alpha$ which produces the best visual results could be calibrated by a standard train-validation-test paradigm.

\section{More details on cubical persistent homology of images}

In this section we provide a more detailed insight into how we use TDA with images in this article. As previously mentioned, for the purpose of assessing shapes we will use the (sublevel set) greyscale filtration in conjunction with cubical peristent homology, which we will henceforth refer to as persistent homology (PH). Before describing persistent homology we must first describe homology.

\subsection{Cubical homology}

Homology is an algebraic method of characterizing notions of connectivity of a shape \citep{edelsbrunner2010}. Though there are various ways to define the homology of an image, we choose to do so in terms of \emph{cubical homology}. Cubical representations of 2-dimensional images most faithfully capture their intuitive shape content \citep{kovalevsky1989}. The main objects of cubical homology are cubical sets. The cubical sets that we consider in this study are collections of unit squares (\emph{2-dimensional elementary cubes}) of the form 
\[
[i, i+1] \times [j, j+1],
\]
along with all intervals (\emph{1-dimensional elementary cubes}) and vertices (\emph{0-dimensional elementary cubes}) on the boundaries, where $i$ and $j$ are integers. If we consider a binary image, then the black pixels and all of the edges and vertices of those pixels in that image constitute a cubical set. Once we have a cubical set associated to a binary image (also known as a \emph{cubical complex}) $X$, we can calculate its homology. 

The most important objects associated to a cubical set $X$ are their homology groups $H_k(X)$, $k = 0, 1, 2, \dots$, which capture $k$-dimensional shape information. The dimensions of these homology groups  are called the \emph{Betti numbers} of $X$ and are denoted $\beta_k(X)$---or $\beta_k$ when $X$ is clear from the context. The $0^{th}$ Betti number $\beta_0$ represents the number of connected components in $X$ and $\beta_1$ represents the number of loops/holes. For more information on cubical homology, one may refer to Chapter 2 of \cite{kaczynski2004}. 

\subsection{Images}

To calculate PH we must define what we mean by an \emph{image}. As in the main document, an $(k \times l)$ image in our setup is most simply taken to be a vector $\tilde{\bm{x}}_i \in \R^{kl}$, where $k$ is the number of rows (\# of vertical pixels) in the image and $l$ is the number of columns (\# of horizontal pixels). If $\bar{\R} = [-\infty, \infty]$ then by considering blocks in the vector $\tilde{\bm{x}}_i$ of length $l$, we may naturally embed $\tilde{\bm{x}}_i$ into $\{f: \mathbb{Z}^2 \to \bar{\R}\}$---identifying some pixel with the origin. Now in $\mathbb{Z}^2$, where spatial information makes sense, it is useful to have another definition at hand for the computation of PH. We follow the lead of \cite{detectda}, in defining a (2-dimensional) \emph{image map} to be a function $I: \mathbb{Z}^2 \to \bar{\R}$, where $I(p) = -\infty$ indicates that $p$ is a black pixel and $I(p)= \infty$ indicates that $p$ is a white pixel. The smallest rectangle with integer coordinates $[k, k+m] \times [l, l+n] \subset \R^2$ which contains all the black pixels---i.e. on which $I < \infty$---will be denoted the \emph{image set} or simply the \emph{image}. As previously mentioned, for the purposes of cubical homology and persistence we must derive cubical sets from the image map $I$. We accomplish this by the construction of another function $I'$ on the family of unit squares with integer vertices. For any such $\tau = [i, i+1] \times [j, j+1]$ we define our filtration function to be
\[
I'(\tau) := I(i,j).
\]
For lower dimensional elementary cubes $\tau$, such as intervals or vertices, we define the value of $I'$ to be the minimum $I(i,j)$ such that $\tau \subset [i, i+1] \times [j, j+1]$. This is consistent with the definition used in the persistent homology software \texttt{GUDHI} Python library, which we use for our PH calculation \citep{gudhi_cubical}. We consider the homology of sublevel set filtrations. That is, we look at the homology of each cubical set 
\[
\{ \tau \equiv [i, i+1] \times [j, j+1]: I'(\tau) \leq t\},
\]
which may naturally be considered as a binary image (each pixel corresponding to a $\tau$). Treating pixels as unit squares (i.e. top-dimensional) rather than elements of $\mathbb{Z}^2$ (i.e. vertices) is also known as the $T$-construction \citep{garin2020duality}. 

\subsection{Persistent homology}

Consider the collection of binary images (i.e. cubical complexes) $\mathcal{I} = \{I_t\}_{t \in \R}$, where 
\[
I_t := \bigcup_{(i,j) \in I^{-1}([0,t])} [i, i+1] \times [j, j+1],
\] 
or equivalently, $I_t =  I'^{-1}([0,t]).$ When $s \leq t$ we have $I_s \subset I_t$ and hence $\mathcal{I} = \{I_t\}_{t \in \R}$ defines a \emph{filtration} of cubical complexes. Given the inclusion maps $\iota_{s,t}: I_s \to I_t$, for $s \leq t$ there exist linear maps between all homology groups
\[
f^{s,t}_k: H_k(I_s) \to H_k(I_t),
\]
which are induced by $\iota_{s,t}$ (see chapter 4 of \citealp{kaczynski2004}). The \emph{persistent homology groups} of the filtered image $\mathcal{I}$ are the quotient vector spaces $\im\, f^{s,t}_k$ whose elements represent shape features---such as connected components or holes---called \emph{cycles} that are ``born'' in or before $I_s$ and that ``die'' after $I_t$. The dimensions of these vector spaces are the \emph{persistent Betti numbers} $\beta_k^{s,t}$. Heuristically, a cycle (more correctly an equivalence class of cycles) $\gamma \in H_k(I_s)$ is born at $I_s$ if it appears for the first time in $H_k(I_s)$---formally, $\gamma \not \in H_k(I_r)$, for $r < s$. The cycle $\gamma \in H_k(I_s)$ dies entering $I_t$ if it merges with an older cycle (born at or before $s$) entering $H_k(I_t)$. The $k^{th}$ persistent homology of $\mathcal{I}$, denoted $PH_k$, is the collection of homology groups $H_k(I_t)$ and maps $f_{s,t}^k$, for $-\infty \leq s \leq t \leq \infty$. All of the information in the persistent homology groups is contained in a multiset in $\R^2$ called the \emph{persistence diagram} \citep{edelsbrunner2010}.  

\begin{figure}
\centering
\includegraphics[width=0.9\textwidth]{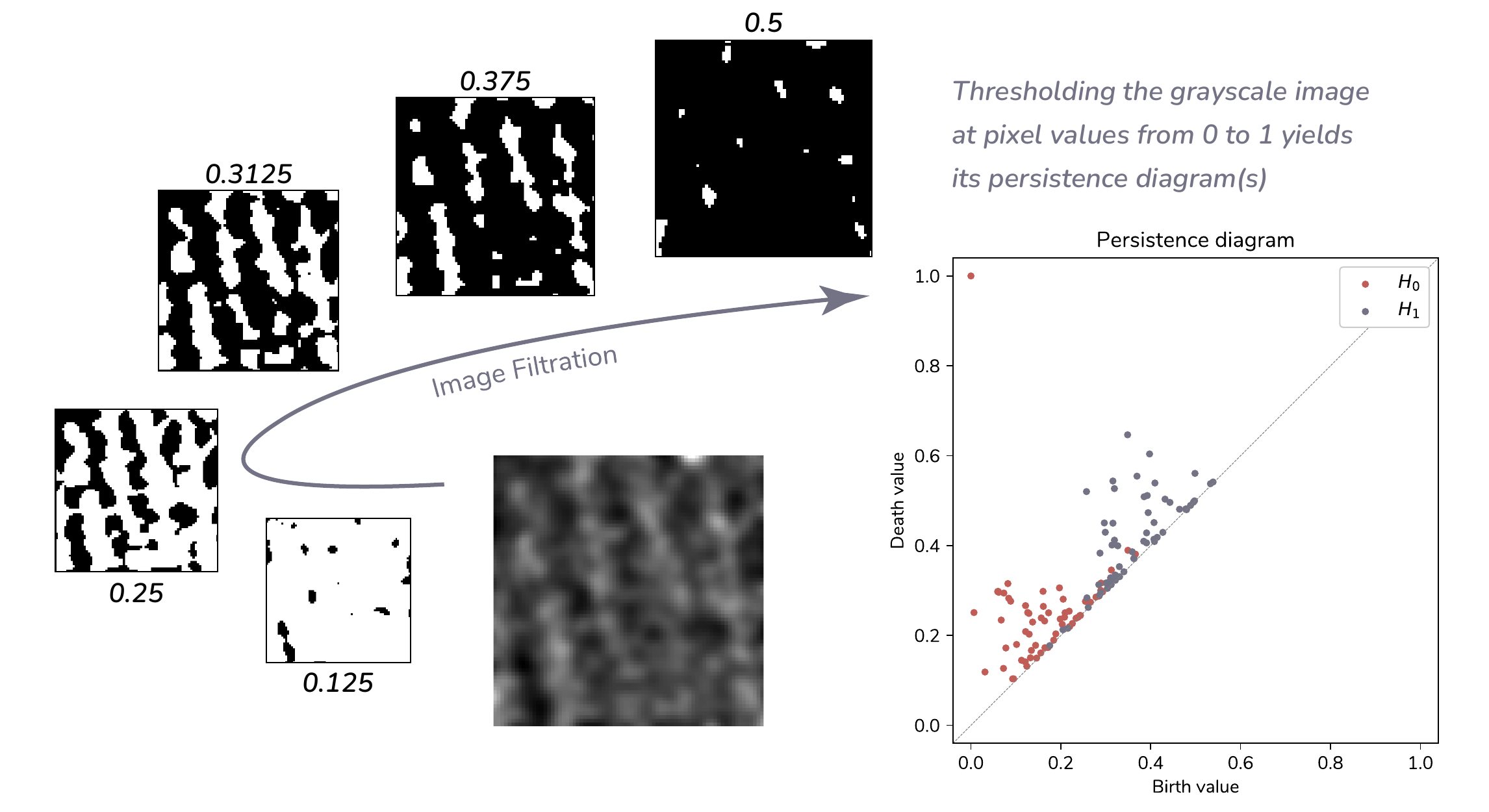}
\caption{An illustration of the 0- and 1-dimensional persistence diagrams of the sublevel set filtration  of a grayscale image. Pixel values at which images were binarized can be seen above the binarized images.}
\label{f:pers_illustration}
\end{figure}
We will denote the $k^{th}$ persistence diagram of $\mathcal{I}$ as $\mathcal{D}^k$. The persistence diagram $\mathcal{D}^k$ consists of the points $(b, d)$ with multiplicity equal to the number of the cycles that are born at $I_b$ and die entering $I_d$. Figure~\ref{f:pers_illustration} contains an illustration of the persistence diagrams associated to a filtration of a given greyscale image. We only consider $\mathcal{D}^0$ and $\mathcal{D}^1$ in this study because higher-dimensional persistence diagrams are trivial for cubical filtrations of 2-dimensional images. For our particular setup, if $(b,d) \in \mathcal{D}^0$, this indicates there is a local minimum of the image $\mathcal{I}$ at some pixel $p^+$ with $I'(p^+) = b$ and $d$ represents the greyscale threshold at which the connected component containing $p$ merges with a connected component containing a local minimum with birth time \emph{less than} $b$. In this case, $p^+$ is called a \emph{positive cell} and gives birth to a connected component in $PH_0$. Furthermore, we can also find an interval $\tau^-$ (negative cell) that kills such a feature, i.e. $I'(\tau^-) = d$ (cf. \citealp{geom_top2018}). An analogous result holds that local maxima kill loops/holes in $PH_1$.

\section{More on the multiple changepoint extension}

Below we discuss the connection between our multiple changepoint method and the multinomial logistic regression setup as well as the results of our multiple BCLR on simulated and real data. 

\subsection{Comparison with multinomial logistic regression setup}

Consider the quasi-likelihood function derived from multinomial logistic regression   
\begin{align*}
\Q(\bmb_1, \dots, \bmb_{J}, \kappa_1, \dots, \kappa_{J} | \bm{X})
=\prod_{j=1}^{J+1} \prod_{i=\kappa_{j-1}+1}^{\kappa_j} \frac{e^{\bm{x}_i^{\top}\bmb_j}}{\sum_{k=1}^{J+1} e^{\bm{x}_i^{\top}\bmb_k}},
\end{align*}
where we define $\kappa_0 \equiv 0$, $\kappa_{J+1} \equiv n$ and $\bmb_{J+1} \equiv 0$. Note also that $0 < \kappa_1 < \kappa_2 < \dots < \kappa_{J} < n$. Let us simplify our notation and denote $\vec{\bmb} = (\bmb_1, \dots, \bmb_{J})$ and $\vec{\kappa} = (\kappa_1, \dots, \kappa_{J})$. We can construct a quasi-posterior density for $(\kvec, \bvec)$ as
$$
\pi(\kvec, \bvec \mid \bm{X}) \propto \Q(\bvec, \kvec \mid \bm{X}) \pi(\bvec, \kvec).
$$
We will assume that $\pi(\bvec, \kvec) = \pi(\bvec)\pi(\kvec)$ and that 
$$
\pi(\kvec) \propto \prod_{j=1}^{J} \alpha_j \ind{ 0 < \kappa_1 < \kappa_2 < \cdots < \kappa_J < n}
$$
for some $\alpha_j > 0$ where $\alpha_j$ may also depend on the index $j-1$. Set $C_{ij} = \log \sum_{k\neq j} e^{\bm{x}_i^{\top}\bmb_k}$ and $\eta_{ij} := \bm{x}_i^{\top}\bmb_j - C_{ij}$. From this, it can be derived that the full conditional for $\kappa_\ell$ is
\begin{equation}\label{e:fc_mbclr}
\pi(\kappa_\ell \mid \kappa_{-\ell}, \vec{\bmb}, \bm{X}) = \frac{\prod_{j=\ell}^{\ell+1} \prod_{i=\kappa_{j-1}+1}^{\kappa_j} \alpha_j(1+e^{-\eta_{ij}})^{-1}}{\sum_{\kappa_{\ell-1} < \kappa_\ell < \kappa_{\ell+1}} \prod_{j=\ell}^{\ell+1} \prod_{i=\kappa_{j-1}+1}^{\kappa_j} \alpha_j(1+e^{-\eta_{ij}})^{-1}},
\end{equation}
where $\kappa_{-\ell} = (\kappa_1, \dots, \kappa_{\ell-1}, \kappa_{\ell+1}, \dots, \kappa_J)$. This means that the full conditional for the estimated changepoint $\kappa_\ell$ only depends on the segment between $\kappa_{\ell-1}$ and $\kappa_{\ell+1}$. Based on this representation, one can think of the proposed multiple changepoint version of BCLR as fixing the segment endpoints $\tau_{\ell-1} \equiv \kappa_{\ell-1}$ and $\tau_{\ell+1} \equiv \kappa_{\ell+1}$ (using the notation from the main document) rather than updating them as one would in a Gibbs sampler. That is, rather than using $\kappa_{\ell-1}^{(t)}$ and $\kappa_{\ell+1}^{(t-1)}$ in \eqref{e:fc_mbclr}, our method simply \emph{doesn't} update our initial values for parameters when sampling $\kappa_{\ell}$. 

Of course, our algorithm can be viewed as a special case of the multinomial logistic regression setup. We ignore \emph{all} the information outside of $(\kappa_{\ell-1}, \kappa_{\ell+1})$, whereas in the multinomial logistic regression setup, the information about all $\bmb_j$ parameters is integrated into \eqref{e:fc_mbclr}. Nonetheless, this note serves to illustrate that it is reasonable to perform inference for changepoints on all consecutive segments $(\kappa_{\ell-1}, \kappa_{\ell+1})$, $\ell = 1, \dots, J$.

\subsection{Performance}

We assessed the performance of the multiple changepoint method described in Section S6 of the main document on a difficult changepoint scenario. In our implementation of this experiment, we used the \texttt{MultiBayesCC} class in the BCLR package mentioned above. We compared the 4 non-parametric methods from the main document: CF, ECP, KCP, and BCLR.

To evaluate the methods, we generated 1000 independent sequences $X_1, \dots, X_{250}$ with
$$
X_i \overset{\text{ind.}}{\sim}
\begin{cases}
N( (0, 0, 0)^{\top}, \Sigma_1) &\mbox{ if }  1 \leq i \leq 100 \\
N( (0, 0, 0)^{\top}, \Sigma_2)  &\mbox{ if }  101 \leq i \leq 175 \\
N( (0, 3, 0)^{\top}, \Sigma_2)  &\mbox{ if }  176 \leq i \leq 205 \\
N( (0, 3, 0)^{\top}, \Sigma_3)  &\mbox{ if }  206 \leq i \leq 250 \\
\end{cases}
$$
where
$$
\Sigma_1 =
\begin{pmatrix}
1 & 0.2 & 0.1 \\
0.2 & 1 & 0 \\
0.1 & 0 & 1
\end{pmatrix}, \hspace{6pt}
\Sigma_2 =
\begin{pmatrix}
4 & 0.4 & 0.2 \\
0.4 & 1 & 0 \\
0.2 & 0 & 1
\end{pmatrix}, \hspace{6pt} \text{ and } \ \ 
\Sigma_3 =
\begin{pmatrix}
4 & 0.4 & -1.7 \\
0.4 & 1 & 0 \\
-1.7 & 0 & 1
\end{pmatrix}.
$$

For CF, we employed the default settings as implemented in \emph{changeforest} package and as described above---though this time we obviously do not limit ourselves to the first changepoint. This means we used 199 permutations for their ``pseudo-permutation'' tests and a ``significance'' level of 0.02. For ECP we also use the default parameters as implemented in the \emph{ecp} package, i.e. $\alpha=1$, 199 permutations and a significance level of 0.05. For KCP we used an RBF kernel, calculating the bandwidth $\gamma$ (where $k(x,y) = e^{-\gamma \lVert x-y\rVert^2})$ via the median heuristic---i.e. $\gamma$ is set to be the inverse of the median of the pairwise distances between observations in a given simulated series. Additionally, the slope heuristic as described in Section 6.2 of \cite{arlot2019} was used for model selection, with $\alpha=2$. We tried KCP with standardized and non-standardized data, as well as a fixed $\gamma = 0.4$, (which was shown to produce good results in \citealp{londschien2022}), but the outcomes did not change very much. Therefore, we only report the best scenario of KCP with the raw data (KCP-raw), using the median and slope heuristics. However, we provided CF, ECP, and BCLR with the standardized, polynomial-embedded data, as this was shown to be beneficial in Section 4.3 of the main document. 

For BCLR we employ two successive warm-up periods utilizing a small number of iterations (discarding the first half of the samples for burn-in, respectively), and removed potential changepoints if their normalized entropy was less than 0.75 and 0.5 respectively. We chose the initial number of segments to be $J = 10$ and gave each $\bmb_j$ a $N(0, I_d)$ prior. After the two warm-up periods, using the resultant partition $\tau_1, \dots, \tau_J$, we fit our multiple changepoint algorithm to the data and the chains on each of the consecutive segments for 5000 iterations. The reason for performing two warm-up periods was to be able to employ all of the data in our final estimation stage, rather than thresholding after having done our ``final round'' of sampling. As with ECP we specified $\Delta = 10$. The thresholds were chosen as they produced reasonable results, but by no means were they calibrated to achieve \emph{optimal} results. The summary of this experiment for all methods can be seen in Table~\ref{t:exp_multi}.

\begin{table}[t]
\centering
\bgroup
\def\arraystretch{1.25}
\begin{tabular}{l|rrrr}
Method & BCLR & CF & ECP & KCP-raw \\
\hline \hline
Rand index & 0.866 & 0.862 & 0.747 & 0.719 \\
& (0.103) & (0.102) & (0.073) & (0.027) \\
\hline
Adjusted Rand index & 0.710 & 0.715 & 0.518 & 0.471 \\
& (0.186) & (0.190) & (0.128) & (0.048) \\
\hline \hline
\end{tabular}
\egroup
\caption{Comparison of various multiple changepoint detection methods in the complicated multiple changepoint scenario described above. Here ``raw'' indicates the method was fed the raw data, rather than the standardized degree-2 polynomial features. Standard deviations for the Rand and Adjusted Rand indices are indicated in parentheses below the mean values.}
\label{t:exp_multi}
\end{table}

Our method performs at least as good as the alternatives, and its variance is comparable to CF (recall that KCP performs slightly \emph{worse} if given the standardized polynomial features). This demonstrates the elite performance of our bottom-up multiple changepoint version of BCLR on a highly difficult task. In \cite{cabrieto2017}, KCP performs well in correlation changepoint tasks but can suffer if certain coordinates (such as $x_2$) do not change correlation structure as well. Our findings here confirm their results. 

Additionally, our method provides much more information than those other methods. For each of the estimated changepoints according to our method there is a ``posterior'' mean $\bmb$ associated to it. As all of the coordinates of our $\x$ series are given standardized to BCLR, the $\bmb$ coordinate with the largest absolute posterior mean should indicate which coordinate of the 9-dimensional embedding
\[
\psi(x) := (x_1, x_2, x_3, x_1^2, x_1x_2, x_1x_3, x_2^2, x_2x_3, x_3^2) 
\]
is driving the change. We shall deem this the ``most prominent coordinate''. Of the 479 instances where BCLR predicted a changepoint less than 5 units from the actual change\footnote{This ensures there will at most one such estimate, owing to $\Delta.$} occurring at index $\kappa^*_1 = 100$\, the $\bmb$ associated to $x_1^2$ (where the change was occurring) had the largest posterior mean 472 times, or 98.5\% percent of those instances. The $x_1^2$ coordinate had either the largest or second largest absolute posterior mean a total of 477/479 times, i.e. 99.6\% of the time. 

Of the 958 instances where BCLR predicted a changepoint within 4 units or less of $\kappa^*_2 = 175$, 431 of those instances had $x_2$ as the most prominent coordinate. Among those cases where $x_2$ was not the most prominent coordinate, $x_2^2$ was the most prominent coordinate 525 out of 527 series, or 99.6\% of the time. Furthermore, BCLR predicted a changepoint in the vicinity (using the same criterion) of $\kappa^*_3 = 205$ a total of 357 times. Of those instances, BCLR identified $x_1x_3$ as the most prominent coordinate 350/357 times, or in 98.0\% of the cases. 

Finally, we analyzed the posterior probabilities for each of the estimated changes in each of the 1000 series. We deem the estimated changepoint with the highest posterior (mode) probability as the ``most prominent change''. Of the 1000 series, 927 of them indicated a most prominent change within 4 units of the true changes at 100, 175, and 205. Of these 927 cases, 872 indicated the most prominent change as index 175, 25 indicated the most prominent change at index 100, and 30 indicated the most prominent change at index 205. 

\subsection{Central England temperature series}

\begin{figure}
\centering
\includegraphics[width=0.75\textwidth]{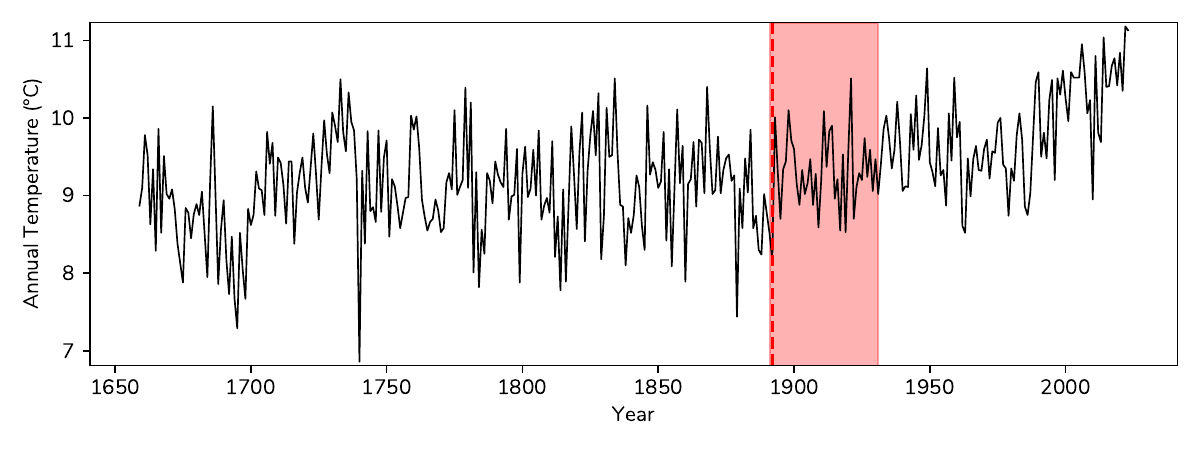}
\includegraphics[width=0.75\textwidth]{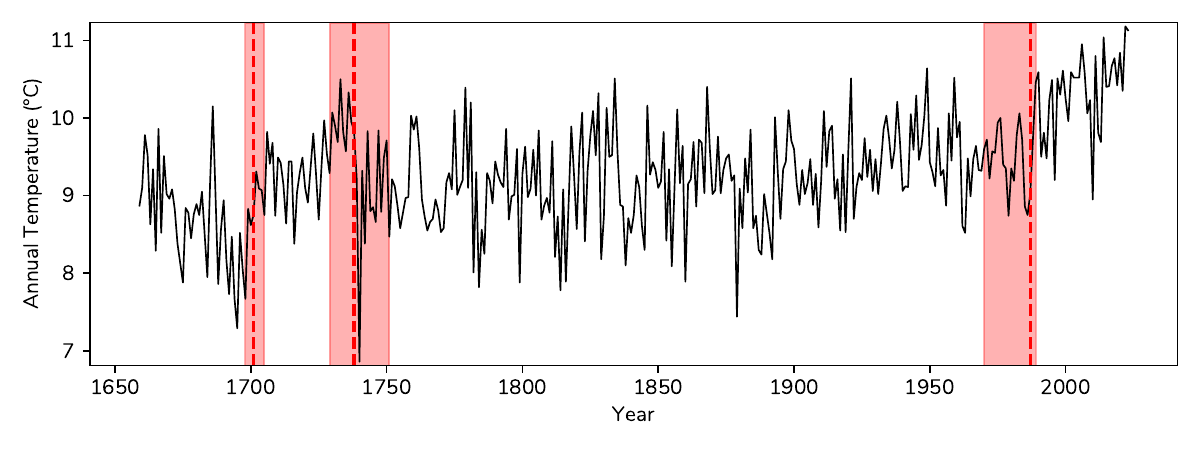}
\caption{Two analyses using BCLR on the CET series. Posterior mode indicated via dashed line and 95\% posterior credible intervals for the estimate change locations indicated in red. The top plot represents an analysis using two warm-up periods with normalized entropy thresholds of 0.75 and 0.5 respectively. The bottom plot represents the output of an analysis where $(\tau_1, \tau_2, \tau_3) = (1700, 1739, 1988)$ and binomial priors are used on each segment. }
\label{f:cet_plots}
\end{figure}

In applying our algorithm as specified in the previous subsection (i.e. the same parameters) to the Central England temperature (CET) series, we observed that BCLR selects only a single changepoint at 1892---see top plot in Figure~\ref{f:cet_plots}. For reference, the analyses considered the ``best'' in \cite{shi2022} flagged changepoints at 1700, 1739 and 1788. These best performing models (in terms of penalized scores) incorporated trend shifts and even temporal dependence into their likelihood functions. 

It is certainly reasonable that there could be a changepoint at 1892 (or at the very least in the 95\% posterior credible interval of $[1891, 1931]$), as this corresponds to a time period characterized by increasing greenhouse gas emissions in the UK and much of the rest of the world \citep{jones2023}. 

If we suppose instead that we would like assess the confidence in the estimates provided in \cite{shi2022}, we could set $(\tau_1, \tau_2, \tau_3) = (1700, 1739, 1988)$ and then run our analysis (without any warm-up period). To encode some degree of belief in these estimates, we specify a binomial prior on each segment with mode at $\tau_i$, $i = 1, 2, 3$. The results of the analysis can be seen in the bottom plot in Figure~\ref{f:cet_plots}. 

The change BCLR seems most confident in is the one estimated to occur at approximately 1700. This is owing to the fact that segment it searches (from 1659 to 1739) shows a high degree of contrast. Further, it is interesting that the estimated change in the late 80's is quantified according to the 95\% credible interval as occurring between 1970 and 1989---where it can be seen a cool period proceeds the relatively warmer decades before warming intensifies beyond 1990. 

Applying CF and KCP to this data yields estimated changepoints of $\{1982, 1988\}$ and $\{1701, 1988\}$ respectively (wherein we use the same settings/parameters from Section S6.2). The changepoints estimated by CF are undesirable, though KCP with the median and slope heuristics produces very reasonable output.  

This is all to say that even though our method does not explicitly incorporate trend or dependence information, it is still capable of producing reasonable results when used off-the-shelf (e.g. top plot of Figure~\ref{f:cet_plots}). It can also provide uncertainty quantification to previous analyses, as in the bottom plot of Figure~\ref{f:cet_plots}. 

\bibliographystyle{plainnat}
\bibliography{ChangepointTDA_V3}